\documentclass[aps,pre,twocolumn,showpacs,showkeys,groupedaddress]{revtex4-1}

\usepackage{verbatim}
\usepackage{graphics}
\usepackage{longtable} 
\usepackage{subfigure}
\usepackage{amsmath}
\usepackage{wrapfig}
\usepackage{epsfig}
\usepackage{float}
\usepackage{amsmath,amssymb,bm}
\usepackage{graphicx}
\usepackage{array}
\usepackage{psfrag}
\usepackage{color}
\usepackage{bbold}

\usepackage{tikz}

\renewcommand{\vec}[1]{\mathbf{#1}}

\newcommand{\figref}[1]{Fig.~\ref{#1}}
\newcommand{\eeqref}[1]{Eq.~\eqref{#1}}

\newcommand{\fh}{\hat{f}}

\newcommand{\CH}[1]{\textcolor{black}{{#1}}}

\begin{document}

\title{Understanding Collective Dynamics of Soft Active Colloids by Binary Scattering}

\author{Timo Hanke, Christoph A.\ Weber and Erwin Frey}

\affiliation{Arnold Sommerfeld Center for Theoretical Physics and Center for NanoScience, Department of Physics, Ludwig-Maximilians-Universit\"at M\"unchen, Theresienstra{\ss}e 37, D-80333 Munich, Germany}

\begin{abstract}
Collective motion in actively propelled particle systems is triggered on the very local scale by nucleation of coherently moving units consisting of just a handful of particles. 
These units grow and merge over time, ending up in a long-range ordered, coherently-moving state.
So far, there exists no bottom-up understanding of how the microscopic dynamics and interactions between the constituents are related to the system's ordering instability.
In this paper, we study a class of models for propelled colloids allowing an explicit treatment of the microscopic details of the collision process. 
Specifically, the model equations are Newtonian equations of motion with separate force terms for particles' driving, dissipation and interaction forces.
Focusing on dilute particle systems, we analyze the binary scattering behavior for these models,
and determine---based on the microscopic dynamics---the corresponding ``collision-rule", \emph{i.e.}, the mapping of pre-collisional velocities and impact parameter on post-collisional velocities.
By studying binary scattering we also find that the considered models for active colloids share the same principle for parallel alignment: the first incoming particle (with respect to the center of collision) is aligned to the second particle as a result of the encounter. This behavior is distinctively different to alignment in non-driven dissipative gases.
Moreover, the obtained collision rule lends itself as a starting point to apply kinetic theory for propelled particle systems in order to determine  the phase boundary to a long-range ordered, coherently-moving state.
 The microscopic origin of the collision rule offers the opportunity to
quantitatively scrutinize the predictions of kinetic theory for propelled particle systems through direct comparison with multi-particle simulations.
We identify local \emph{precursor} correlations at the onset of collective motion to constitute the essential determinant 
for a qualitative and quantitative validity of kinetic theory.
In conclusion, our ``renormalized" approach  clearly indicates that the framework of kinetic theory is flexible enough to accommodate the complex behavior of soft active colloids
and allows a bottom-up understanding of how the microscopic dynamics of binary collisions 
relates to the system's behavior on large length and time-scales.
\end{abstract}



\maketitle


\section{Introduction}\label{sect:introduction}

The emergence of collective motion in active systems composed of (self-)propelled entities is a highly complex self-organization phenomenon~\cite{Aranson_Tsimring_book, Julicher_gel_review, Frey_Review2008, Ramaswamy_Review, Vicsek_Review, Zia_Review2011, Marchetti:2012ws}.
Examples include systems on vastly different length scales, ranging from micrometer-sized systems, made of biological constituents~\cite{Butt, Dombrowski_2004,Schaller,Yutaka,Schaller2,Ringe_PNAS,pnas_bacteria,Dogic,howard2012}, over millimeter-sized vibrating hard granules~\cite{Dauchot_Chate_2010, Dauchot_long, Kudrolli_2008},  to large cooperative animal groups~\cite{Couzin200936, ballerini_starlings, Lopez06122012}.
 While all these systems share the capacity to form collective patterns, the precise mechanisms how they are formed and their degree of universality remain elusive.

Historically, generic agent-based models~\cite{Vicsek,Czirok_2000,Chate_2004} constitute the first theoretical approach aiming to understand the minimal ingredients necessary for the emergence of collective motion. In the Vicsek model~\cite{Vicsek} collective motion is thought to be a consequence of a generic competition between a local tendency of ``ferromagnetic parallel alignment" and noise. Specifically, particle alignment is implemented as an update rule in the spirit of a cellular automaton: Each particle aligns parallel to the average of all particles' orientations within some defined finite neighborhood. Vicsek-like models have been instrumental in exploring the pattern forming capabilities of active systems~\cite{Vicsek,Czirok_2000,Chate_2004,Chate_Variations, Chate_long, Albano2009}. 
Nevertheless, there are also drawbacks of such generic agent-based models.  Most importantly, they do not account for the physics of the interaction between active constituent particles. In the meantime there have emerged however a range of well characterized experimental model systems including actin and microtubule gliding assays~\cite{Schaller,Yutaka}, and shaken granular particles~\cite{Dauchot_long,VibratedDisks2013},  which are amenable to a highly quantitative analysis at the scale of collisions between individual particles. 

These microscopic features can, in principle, be analyzed in terms of molecular dynamics simulations: While in cellular automata like the Vicsek model particles move at constant speed, Newtonian dynamics explicitly account for driving forces and dissipation. Moreover, instead of update rules there are interaction forces between the constituent particles. The ensuing equations of motion allow for a more realistic description of individual particle trajectories as well as resolving the time and length scales of pairwise collisions. Although previous studies of such models show an even richer spatio-temporal dynamics than Vicsek-like models~\cite{Levine_2000, Erdmann_2005, Orsogna_2006, Mach_daphina_2007, Grossman_2008, Erdmann_Ebeling_2000, pawel_2011, grossmann_pawel_2012, pawel_longpaper_2012, pawel_locusts_2009}, a thorough analysis of how order builds up from microscopic particle interactions has never been undertaken. Here we try to close this gap and ask, focusing on dilute conditions: Can we understand the collective dynamics of active particle systems by solely considering \emph{binary} interactions between the constituents? 

We address this question by a combination of molecular dynamics (MD) simulations and kinetic theory. Specifically, we consider a system of active spherical particles interacting by a short-ranged and repulsive harmonic force. In the absence of interactions, these soft particles move at constant speed determined by a balance between a driving and a dissipative force. From MD simulations for this system of \emph{soft active colloids} we determine the phase boundary between the isotropic and the polarized state. To connect these numerical studies with a kinetic approach, we employ the MD simulations to also analyze binary scattering of particles. Thereby one can extract the underlying collision-rule, \emph{i.e.}, the mapping of the pre-collisional to the post-collisional velocities, which lends itself as a starting point to set up a Boltzmann equation for the one-particle distribution function. Following previous approaches in the literature \cite{Aronson_MT, Bertin_short, Bertin_long}, the latter can also be used to determine the phase boundary. Any mismatch between this phase boundary and the one obtained from the MD simulations must be directly related to the assumptions underlying the Boltzmann equation. Since these assumptions concern the nature of the particle collisions and the ensuing correlations between the positions and orientations of the particles, a quantification of the mismatch will allow a deepened understanding of the ordering process in active systems and highlight the differences to thermodynamic systems. In particular, it will be interesting to see to what degree the molecular chaos assumption of classical Boltzmann's theory remains valid for active systems. 

The structure of this paper is the following: 
The models for soft active colloids are described in detail in Sect.~\ref{sect:models}.
We present the phase boundary between the isotropic and polarized states obtained from the molecular dynamics simulations in section~\ref{sect:multi_particle_simulation}.
 Section~\ref{sect:binary_scattering} is devoted to the MD study of  two particle interactions: the binary scattering study. 
 Specifically, the collision geometry is introduced in Sect.~\ref{sect:collision_geometry} and the parameters characterizing the strength of alignment are described in Sect.~\ref{sect:alignment_strength}. 
Then, we discuss the results of the binary scattering study in Sect.~\ref{sect:scattering_study}. 
Section~\ref{sect:collective} deals with the collective dynamics of one of the models for soft active colloids. 
Specifically, we explain in Sect.~\ref{sect:collective_coarse_grained_col_rule} how the collision rule is implemented into the kinetic description, and set up the corresponding Boltzmann equation in Sect.~\ref{sect:collective_kinetic_model}.
Finally, in Sect.~\ref{sect:collective_comparison}, we derive the phase boundary between the isotropic and polarized state from the Boltzmann equation and 
compare it to the phase boundary obtained from the MD simulations.  In this section we also show how the kinetic approach has to be modified in order to account for correlations close to the phase transition to collective motion (termed as \emph{precursor} correlations). We close by a concise conclusion and outlook in the final Sect.~\ref{sect:conclusion_outlook}.

\section{The Dynamic Models}\label{sect:models}

\subsection{Deterministic equations of motion}\label{sect:models_det}

We study \emph{dynamic models} for active colloids in two dimensions \cite{Levine_2000,Erdmann_2005,Orsogna_2006,Mach_daphina_2007,Grossman_2008} in terms of Newtonian equations of motion including the following forces:  (i) An active propelling force $F^{\text{prop}}_i \vec{\hat{v}}_i$ [with $\vec{\hat{v}}_i= \vec{v}_i / \left|\vec{v}_i\right|$: unit vector of the velocity] capturing the internal propulsion mechanism that in turn is balanced by (ii) a dissipative force $-F^{\text{diss}}_i \vec{\hat{v}}_i$ accounting for the particle's loss of kinetic energy.  Finally, (iii) particles interact by means of a two-body interaction force denoted as $\vec F_{ij}$, which may be attractive, repulsive, or any combination thereof. The ensuing Newtonian equations of motion read
\begin{equation} \label{eq:constitutive_general}\
 \frac{\text{d}}{\text{d}t} \, \vec{v}_i = \left( F^{\text{prop}}_i  - F^{\text{diss}}_i  \right) \vec{\hat{v}}_i + \sum_j \, \vec F_{ij},
\end{equation}
where we use units such that the mass of each particle is set to unity. The propelling and dissipative forces are commonly taken as~\cite{Levine_2000,Erdmann_2005,Orsogna_2006,Mach_daphina_2007}
\begin{subequations}
\begin{align} 
F^{\text{prop}}_i \left(\left| \vec{v}_i \right|\right) &= \alpha \, {\left| \vec{v}_i \right|}^\nu, \\
F^{\text{diss}}_i \left(\left| \vec{v}_i \right|\right) &= \beta \, {\left|\vec{v}_i\right|}^\gamma,
\end{align}
\end{subequations} 
with  exponents $\nu$ and $\gamma$ characterizing each force's respective dependence on the particle velocity.
Their choice must fulfill the condition $\gamma > \nu$ to ensure a proper definition of a (stable) stationary velocity $v_0={\left(\alpha/\beta\right)}^{1/(\gamma-\nu)}$. 
To accommodate a more complex dependence on $\vec{v}_i$, the amplitudes $\alpha$ and $\beta$ may in general be functions of the particle's velocity \cite{Mach_daphina_2007,Grossman_2008}.

The interaction force $\vec{F}_{ij}$ exerted by particle $j$ on particle $i$ is a function of the relative position $\vec{r}_i - \vec{r}_j$ \cite{Levine_2000,Erdmann_2005,Orsogna_2006,Mach_daphina_2007}, with $\vec{r}_i$ denoting the position of particle $i$. In addition, it may depend on the particles' velocities, accounting for inelastic interactions between the particles \cite{Grossman_2008}. Here, we restrict ourselves to short-ranged repulsive interactions between the particles: 
If two particles $i$ and $j$ of radius $R$ exhibit some finite overlap, $\xi_{ij} = \left(2R - \left|\vec{r}_i - \vec{r}_j\right|\right) > 0$, there is a harmonic repelling force $\vec{F}_{ij}= K \, \xi_{ij} \, \vec{\hat{r}}_{ij}$ acting on particle $i$ in the direction of $\vec{\hat{r}}_{ij}={(\vec{r}_i-\vec{r}_j)}/{\left|\vec{r}_i-\vec{r}_j\right|}$. The coefficient $K$ denotes the stiffness of the harmonic spring.
\CH{Note that the choice of a harmonic interaction is made for simplicity and our qualitative results do not depend on the specific form of the interaction potential.}

In the following, we focus on two specific \emph{dynamic models}, where $\alpha$ and $\beta$ are constants. In the first model, hereafter referred to as \emph{model A}, there is a constant propulsion in the direction of motion, while the dissipation has a linear dependence on the particle's velocity ($\nu=0$, $\gamma=1$).  The corresponding equations of motion are then given by \cite{Levine_2000, Erdmann_Ebeling_2000, pawel_2011, pawel_longpaper_2012}:
\begin{equation} \label{eq:constitutive_model_A}
 \frac{\text{d}}{\text{d}t}\, \vec{v}_i = \left(\alpha - \beta \, \left|\vec{v}_i\right| \right) \vec{\hat{v}}_i + \sum_j \, \begin{cases} 
K \, \xi_{ij} \, \vec{\hat{r}}_{ij}   &\mbox{if }    \xi_{ij}\ge 0 ,\\ 
0  & \mbox{else} .
\end{cases}
\end{equation}
The second model considered, referred to as \emph{model B}, features a driving force that scales linearly with the particle's velocity, and a dissipative force that is cubic in the velocity ($\nu=1$, $\gamma=3$) \cite{Erdmann_2005, Orsogna_2006}:
\begin{equation} \label{eq:constitutive_model_B}
 \frac{\text{d}}{\text{d}t}\, \vec{v}_i = \left(\alpha \, \left|\vec{v}_i\right| - \beta \, {\left| \vec{v}_i \right|}^3 \right) \vec{\hat{v}}_i + \sum_j \,  
\begin{cases} 
K \, \xi_{ij} \, \vec{\hat{r}}_{ij}   &\mbox{if }    \xi_{ij}\ge 0 ,\\ 
0  & \mbox{else} .
\end{cases}
\end{equation}

\subsection{Rescaled model equations}\label{sect:rescaled_equations}

\subsubsection*{Model A}
For deterministic particle motion governed by \eeqref{eq:constitutive_model_A} in the absence of interactions, the direction of the velocity does not change, while the particle's speed $v(t) = \left|\vec{v}(t)\right|$ exponentially approaches the stationary value $v_0 = \alpha/\beta$ as $v(t)=v_0 + \big(v(t_0)-v_0\big) e^{-\beta(t-t_0)}$ \CH{\footnote{\CH{In the marginal case $v(t_0)=0$ the exponential approach to $v_0$ is absent since $\vec{\hat{v}}$ is undefined. The particle never experiences an accelerating force and therefore $v(t)=0\ \forall t$.}}}, with a characteristic relaxation time 
\begin{eqnarray}
\tau_\text{eq} = \beta^{-1} \, .
\end{eqnarray}
Rescaling the particles' spatial coordinate $\vec{\tilde{r}}_i=\vec{r}_i/d$ by the particle diameter $d=2R$ and time by the relaxation time $\tau= t/\tau_\text{eq} = \beta t$, we arrive at the following dimensionless equations:
\begin{equation} \label{eq:rescaled_model_A}
  \frac{\text{d}}{\text{d}\tau}\, \vec{\tilde{v}}_i = \big( \mu - \left|\vec{\tilde{v}}_i\right| \big) \vec{\hat{v}}_i +  \sum_j  \begin{cases} 
\kappa \,\tilde{\xi}_{ij} \, \vec{\hat{r}}_{ij}   &\mbox{if }   \tilde \xi_{ij}\ge 0 ,\\ 
0  & \mbox{else} ,
\end{cases}
\end{equation}
where $\vec{\tilde{v}}_i = \text{d}\vec{\tilde{r}}_i / \text{d}\tau$ is the dimensionless velocity and  $\tilde{\xi}_{ij}= \xi_{ij}/d$ is the rescaled penetration depth. The rescaled interaction constant $\kappa={K}/{\beta^2}$, and the rescaled stationary speed $\mu={\alpha}/{(\beta^2 d)}$ constitute the two key parameters of the model.

For the short-ranged harmonic interaction potential, the duration of a particle encounter is of the order of $\tau_\text{int}:=K^{-1/2}$. Hence, the dimensionless parameter $\kappa$ compares the characteristic relaxation time of the velocity $\tau_\text{eq}= \beta^{-1}$ to the duration of a particle interaction, and can be rewritten as 
\begin{equation} \label{eq:def_kappa}
	\kappa = \left( \frac{\tau_\text{eq}} {\tau_\text{int}} \right)^2 .
\end{equation}
Therefore, $\kappa \gg 1$ signifies that the repulsive interaction force dominates the collision, whereas for $\kappa \ll 1$ the dissipative force is the major factor governing the dynamics during collisions.
In the following we will refer to $\kappa$ as the \emph{interaction parameter}.

The dimensionless parameter $\mu$ can be interpreted as follows: 
Consider the length $l_\text{eq}=v_0 \, \tau_\text{eq}$ that a particle moving with the stationary speed $v_0=\alpha/\beta$ covers within a time interval equal to the relaxation time $\tau_\text{eq} = \beta^{-1}$. The ratio of this length scale $l_\text{eq}$ and the particle diameter $d$ then is equal to the parameter $\mu$: 
\begin{equation} \label{eq:def_mu}
	\mu= \frac{l_\text{eq}}{d}.
\end{equation} 
Since $l_\text{eq}$ provides an estimate for the relaxation length in the system, \emph{i.e.}, the distance traveled by a particle until its velocity is relaxed,  small values of $\mu$ signify that particles are moving in a highly damped system, while for high values of $\mu$ damping is weak. Therefore, we will hereafter refer to $\mu$ as the dimensionless relaxation length or in short \emph{relaxation length}.

\subsubsection*{Model B}
In the absence of interactions, the speed of particles relaxes according to \eeqref{eq:constitutive_model_B} \CH{from any non-zero value $v(t_0)$} to the stationary speed 
$v_0 = \left({\alpha/\beta}\right)^{1/2}$ as 
$$v(t)=\left[ \frac{1}{{v_0}^2} + \left( \frac{1}{{v(t_0)}^2} - \frac{1}{{v_0}^2} \right) e^{-2\alpha(t-t_0)} \right]^{-1/2},$$
 with a characteristic relaxation time $\tau_\text{eq} = \alpha^{-1}$ instead of $\beta^{-1}$ as for \emph{model A}. Again, rescaling length by particle diameter $\vec{\tilde{r}}_i=\vec{r}_i/d$ and time by the relaxation time $\tau= t/\tau_\text{eq}$ gives the following dimensionless equations for \emph{model~B}:
\begin{equation} \label{eq:rescaled_model_B}
  \frac{\text{d}}{\text{d}\tau}\, \vec{\tilde{v}}_i = \left( 1 - \frac{{\left|\vec{\tilde{v}}_i\right|}^2}{\mu^2} \right) \vec{\tilde{v}}_i +  \sum_j 
\begin{cases} 
\kappa \,\tilde \xi_{ij} \, \vec{\hat{r}}_{ij}   &\mbox{if }   \tilde \xi_{ij}\ge 0 ,\\ 
0  & \mbox{else} ,
\end{cases}
\end{equation}
where $\vec{\tilde{v}}_i = \text{d}\vec{\tilde{r}}_i / \text{d}\tau$ is the dimensionless velocity, and  $\tilde \xi_{ij}= \xi_{ij}/d$ is the rescaled penetration distance.  

The two parameters of the model are given by $\protect{\kappa = K/\alpha^2}$ and $\protect{\mu = {\left(d^2 \alpha \beta \right)}^{-1/2}}$. 
The interpretation of $\kappa$ as the squared ratio of the relaxation to the interaction timescale, and of $\mu$ as the ratio of the relaxation length $v_0\tau_\text{eq}$ to the particle diameter is similar to the corresponding parameters for \emph{model A}. 
This allows the comparison of the alignment capabilities for the two models arising from the difference in the form of the equations alone. In the following, we will not distinguish between the parameters $\kappa$ and $\mu$ of \emph{model A} and \emph{model B}, respectively.

\subsection{Description of random fluctuations}\label{sect:models_noise}

For the study of systems consisting of a large number of particles, the dimensionless, deterministic equations of motion [\eeqref{eq:rescaled_model_A} or \eeqref{eq:rescaled_model_B}] are complemented by a stochastic element accounting for noise in the system. Since Brownian noise is  irrelevant in most active systems~\cite{Schaller,Kudrolli_2008,Dombrowski_2004,pnas_bacteria,Yutaka,Dauchot_Chate_2010,Dauchot_long,VibratedDisks2013}, we restrict ourselves to a stochastic element that solely leads to fluctuations in the particles' orientations; see \emph{e.g.}\ Refs.~\cite{Vicsek,Chate_2004,Chate_long,Grossman_2008}. Specifically, we implement noise as an additive, uncorrelated stochastic force 
periodically changing the particles' orientation with a frequency $\lambda$. Denoting the velocity at time $t$ obtained from integrating the deterministic model equations as $\vec{\hat{v}}_i^{(\text{det})}(t)$,  this direction is rotated by a Gaussian-distributed random angle $\eta$ leading to the stochastic velocity direction $\vec{\hat{v}}_i(t)$:
\begin{equation} \label{eq:def_angular_noise}
	\vec{\hat{v}}_i(t) = 
	\begin{pmatrix} 
		\cos\eta & -\sin\eta \\
		\sin\eta & \cos\eta
	\end{pmatrix}
	\vec{\hat{v}}_i^{(\text{det})}(t) .
\end{equation} 
The Gaussian distribution of the random angle $\eta$ has zero mean and variance $\sigma_0^2$. 
In general, the parameters $\lambda$ and $\sigma_0$ together determine the strength of noise in the system.
However, for all our later studies the frequency  $\lambda$ is not a central issue.
As a matter of convenience, we choose the time between two stochastic events equal to the discrete time step $\Delta t$ used for the numerical integration of the equations of motion, \emph{i.e.}, $\lambda^{-1}/\tau_\text{eq}=\Delta t/\tau_\text{eq}=0.1$~\CH{\footnote{\CH{Please note when varying the time discretization $\Delta t$, the noise strength $\sigma_0$ has to be adjusted according to $\sigma_0 \sim \Delta t^{0.5}$ in order to avoid a diverging noise level for $\Delta t \to 0$.}}.}
This leaves us with $\sigma_0$ to determine the strength of noise in the multi-particle simulations presented in the next section.

\section{Multi-particle simulations}\label{sect:multi_particle_simulation}

As a reference point for our subsequent bottom-up analysis of the emergence of collective motion, we performed molecular dynamics simulations of a large number of particles to determine the phase diagram. 
To this end, the deterministic equations of motion for \emph{model A} [\eeqref{eq:rescaled_model_A}], supplemented by the angular noise described in \eeqref{eq:def_angular_noise}, were integrated numerically.
For the parameters of the model, we chose $\mu=0.05$ and $\kappa=0.1$
\footnote{This choice of parameter values is justified in retrospect by the results detailed in section~\ref{sect:scattering_study}: The employed values correspond to a maximal alignment capability of \emph{model A} obtained from analyzing the scattering of two particles.}. 
Particles moved in a square box of linear size $L=250d$ with periodic boundary conditions. We considered particle numbers in the range $N\sim10^5$--$10^6$.
As initial configuration particles were placed randomly in the simulation box. 
Overlapping particles were relocated until any particles' overlap had vanished 
 \footnote{For larger packing fractions, this procedure becomes unfeasible. To complete the numerical phase diagram, starting from random positions we used an over-damped algorithm prior to the actual simulation, where only the interaction forces induce movement until remaining overlaps have been minimized.}. 
Particle velocities were given by randomized directions, with their modulus set equal to the stationary velocity, which is given by $\mu$ in dimensionless units.

\begin{figure}[tb]
\begin{center}
\includegraphics[width=\linewidth]{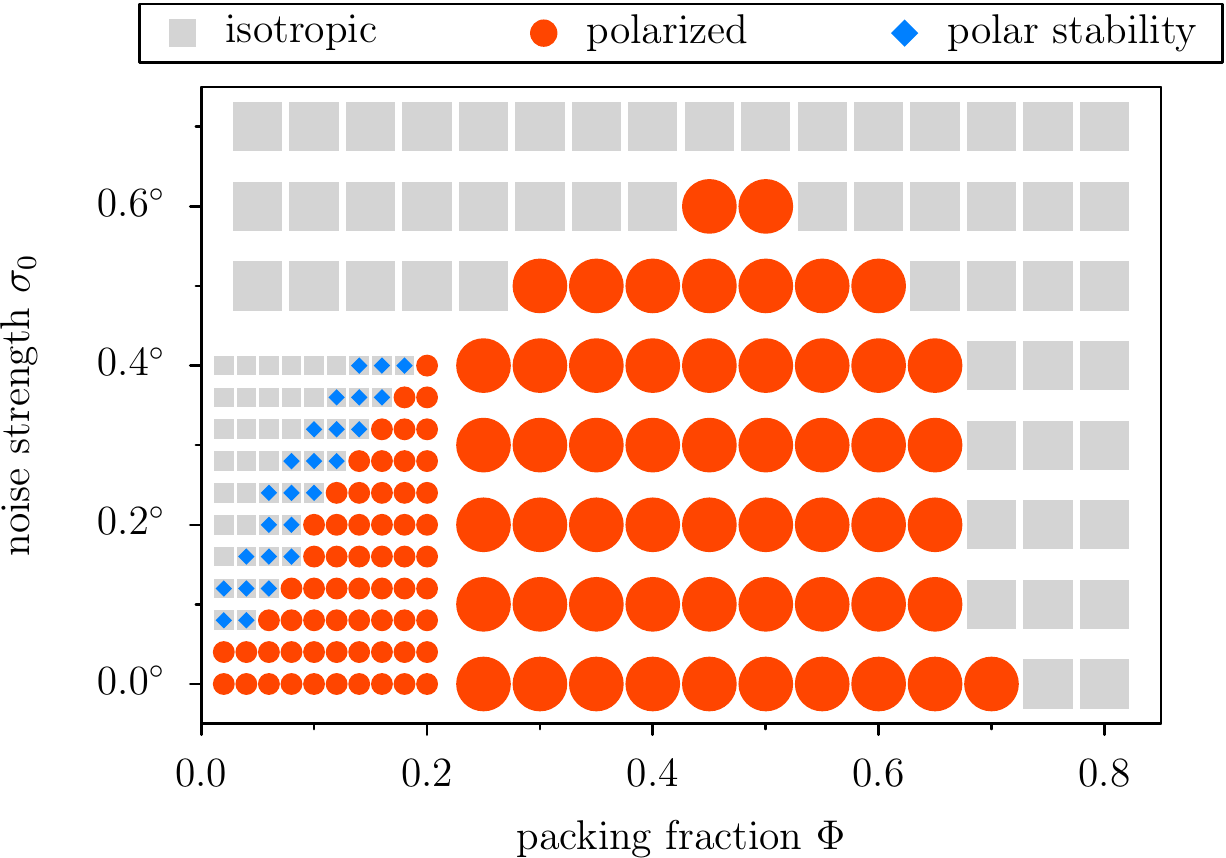}
\caption{
(color online)
\emph{Phase diagram} as function of packing fraction $\Phi$ and angular noise strength $\sigma_0$ [in degrees] obtained from multi-particle simulations. 
Red dots indicate polarized states while grey squares correspond to control parameter sets where no change in the initial isotropic state was observed. 
Initially perfectly polar states ($\psi=1$) that remained polarized over the entire simulation time are marked by blue diamonds.}
\label{plot:phase_diagram_simulation}
\end{center}
\end{figure}

In order to numerically determine the phase boundary, we computed typically $10$ realizations of different initial coordinates and velocity directions for a set of values of the single particle noise $\sigma_0$ and the packing fraction 
\begin{equation}
	\Phi=\frac{N\pi d^2}{4L^2}. 
\end{equation}
Running the simulations for sufficiently long times, we classify a point in  $\Phi$-$\sigma_0$-parameter space to be macroscopically \emph{polarized} if the system's polarization
\begin{equation}  
	\psi=\frac{1}{N} \sum_{i=1}^{N} \hat{\vec{v}}_i,
\end{equation}
 given as the average over all particles' velocity directions $\hat{\vec{v}}_i=\vec{v}_i/ \left|\vec{v}_i\right|$, exceeded a value of $0.6$ for at least one realization.
Otherwise, the parameter set is termed as \emph{isotropic}.
The resulting phase diagram is depicted in \figref{plot:phase_diagram_simulation}: 
Red dots indicate the values of the control parameters ($\Phi, \sigma_0$) where a transition to a polar state was observed, while grey squares correspond to isotropic states.
 Close to the phase boundary the dynamics exhibits critical slowing down, \emph{i.e.}, the time until the system builds up a polar state grows quickly beyond computationally accessible time-scales. Therefore, for small packing fractions $\Phi$ corresponding to small critical noise strengths $\sigma_{0,\text{c}}$, we prepared multiple realizations of systems in a perfectly polar state ($\psi=1$, but with randomized spatial initial configurations), and studied whether this polar state remains stable or disintegrates into an isotropic state (only investigated for $(\Phi,\sigma_0)\in[0,0.2]\times[0^\circ,0.4^\circ]$). Those states where all realizations remained polarized throughout the simulation time interval of $10^6$ time units are indicated by blue diamonds in \figref{plot:phase_diagram_simulation}.
 
For packing fractions $\Phi\le0.5$, the phase diagram shows the shape expected from the generic picture of a competition between aligning interactions and noise~\cite{Vicsek,Chate_2004,Chate_Variations, Chate_long}:  
The critical noise strength $\sigma_{0,\text{c}}$, at which a transition from an isotropic to a polarized state occurs, shows a monotonic increase with the value of the packing fraction $\Phi$. 
Additionally, we find hysteresis as observed in the Vicsek model~\cite{Chate_long}: 
There is a small range of control parameters ($\Phi, \sigma_0$) adjacent to the phase boundary where polar states are always stable, but 
polar ordering from an isotropic state could not be observed.
Finally, in the regime of large packing fractions polarized states do not develop from initial isotropic states, which has also been found in other agent-based models~\cite{Menzel_Ohta_2012}.

\section{Binary Scattering Study}\label{sect:binary_scattering}

In the previous section we have shown that polar order can emerge from an unordered, homogeneous state
for large enough packing fraction $\Phi$ and small enough angular noise strength $\sigma_0$.
For dilute conditions ($\Phi\ll1$), one expects that binary particle interactions are the dominant process leading to the alignment of particles. Therefore, we performed binary scattering studies using the deterministic equations of motion given in section~\ref{sect:models_det}, and analyzed how the interplay between driving, dissipation and the strength of the repulsive  interaction affects the ordering propensity of the active system.

Specifically, the aim of the scattering study is two-fold: 
(i)~First, we systematically compare models with different driving and dissipation forces regarding their capabilities to induce alignment through binary collisions. In particular, we will work out conditions for optimal alignment, and compare the underlying principles leading to parallel alignment. (ii)~Second, we determine the ``collision-rule,'' \emph{i.e.}, the mapping of pre-collision velocities and impact parameter onto post-collisional velocities. This mapping then constitutes the foundation for a mesoscopic description of the system in the framework of kinetic theory, detailed in section~\ref{sect:collective}. Due to its microscopic origin the collision rule allows to scrutinize the predictions of kinetic theory for propelled particle systems through direct comparison with the multi-particle simulations presented in section~\ref{sect:multi_particle_simulation}.

\subsection{Collision geometry}\label{sect:collision_geometry}

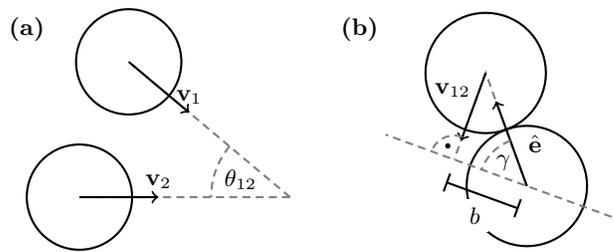
\begin{figure}[tbrl]
\begin{center}
\begin{minipage}[H]{0.49\linewidth}
\begin{tikzpicture}
	[scale=0.7, dashedlines/.style={thick,color=gray,densely dashed}, lines/.style={thick,color=black}, circles/.style={thick,color=black},]
	\node at (-5,3.2) {\textbf{(a)}};
	\draw[circles] (0,0) node[coordinate] (zero) {} +(180:4) circle (1) node[coordinate] (two) {} +(140:4) circle(1) node[coordinate] (one) {};
	\draw[dashedlines] (zero) -- (two) (zero) -- (one);
	\draw[lines,->] (two) -- +(0:1.5) node[above] {$\vec{v}_{2}$};
	\draw[lines,->] (one) -- +(-40:1.5) node[above] {$\vec{v}_{1}$};
	\draw[dashedlines] (zero) ++(140:1.5) arc(140:180:1.5);
	\draw (zero) +(160:1) node {$\theta_{12}$};
\end{tikzpicture}
\end{minipage}
\begin{minipage}[H]{0.49\linewidth}
\begin{tikzpicture}
	[scale=0.8, rotate=-20, dashedlines/.style={thick,color=gray,densely dashed}, lines/.style={thick,color=black}, circles/.style={thick,color=black}]
	\node at (-3.5,1.5) {\textbf{(b)}};
	\draw[circles] (0,0) circle (1) node[coordinate] (two) {} +(130:2) circle(1) node[coordinate] (one) {};
	\draw[dashedlines] (two) +(180:-1.5) -- +(180:2.5) +(0,0) -- (one) -- +(-90:2*0.766) node[coordinate] (b) {};
	\draw[lines,->] (two) -- node[right=3pt] {$\hat{\vec{e}}$} +(130:1.5) ;
	\draw[lines,->] (one) -- node[left,near start] {$\vec{v}_{12}$} +(-90:1.2) ;
	\draw[dashedlines] (b) +(90:0.4) arc(90:180:0.4) ;
	\draw[fill=black] (b) +(135:0.25) circle(1pt);
	\draw[dashedlines] (two) ++(130:0.8) arc(130:180:0.8) ;
	\draw (two) +(155:0.55) node {$\gamma$};
	\path (two) +(0,-0.4) node[coordinate] (startb) {} (b) +(0,-0.4) node[coordinate] (endb) {};
	\draw[lines,|-|] (startb) -- node[fill=white,below=2pt,pos=0.6] {$b$} (endb);
\end{tikzpicture}
\end{minipage}
\caption{Illustration of the collision geometry defined by the impact parameter $b$ and the relative angle $\theta_{12}$. The point of intersection of the particles' orientations, indicated by the dashed lines, defines the \emph{center of collision}.  \textbf{(a)} Particles $1$ and $2$  moving with their respective velocities $\vec{v}_{1}$ and $\vec{v}_{2}$, enclosing the relative angle $\theta_{12}=\protect\angle{\left(\vec{v}_{1}, \vec{v}_{2}\right)}$. \textbf{(b)} Definition of  relative velocity $\vec{v}_{12}$, unit vector $\vec{\hat{e}}$, and impact parameter $b$. The unit vector $\vec{\hat{e}}$ is defined as the normalized  vector connecting the particle centers at the very moment of contact, where $b=-2R\cos\gamma$ gives the offset from a head-on collision.}
\label{pic:coll_geometry}
\end{center}
\end{figure}

Due to the short-ranged nature of the repulsive interaction potential [in~\eeqref{eq:rescaled_model_A} and~\eeqref{eq:rescaled_model_B}]
one can give a precise definition of the instant when two particles come into contact [\figref{pic:coll_geometry}(a)]. 
Capturing all possible configurations for particle encounters then amounts to defining an appropriate set of parameters describing the geometry at this first moment of contact. Denoting the particles' positions as $\vec{r}_1$ and $\vec{r}_2$, the inter-particle distance is $\left|\vec{r}_1-\vec{r}_2\right|=2R$ at contact, and the spatial arrangement 
of the collision is appropriately described by the unit vector $\vec{\hat{e}}$ that defines the normal direction~\cite{Brilliantov_book}:
\begin{equation}\label{eq:def_e}
	\vec{\hat{e}} = \frac{ \vec{r}_1 - \vec{r}_2 }{ \left| \vec{r}_1 - \vec{r}_2 \right| } .
\end{equation}
In combination with the relative velocity $\vec{v}_{12} = \vec{v}_1 - \vec{v}_2$, the unit vector $\vec{\hat{e}}$ completely determines the geometry of the collision at the moment of contact. 
Instead of these vectorial quantities, however, it is more convenient to work with two equivalent scalar parameters.
Since the relative position of the particles only matters with respect to the direction of their relative velocity, we introduce the angle $\gamma$ as a parametrization for the unit vector $\vec{\hat{e}}$ [\figref{pic:coll_geometry}(b)],
\begin{equation}
	 \sin  \gamma = - \frac{ \vec{v}_{12} \cdot \hat{\vec{e} }}{ \left| \vec{v}_{12} \right| } .
\end{equation}
In our dimensionless units where the particle diameter $d$ constitutes the unit length, the impact parameter $b$ is then defined as \cite{Brilliantov_book}
\begin{equation} \label{eq:def_impact} 
 	b = - \cos  \gamma  .
\end{equation}
The impact parameter characterizes the type of collision: $b=0$ signifies a head-on collision in the relative frame or symmetric collision in the laboratory frame, whereas $b=\pm1$ corresponds to glancing collisions where particles are barely touching each other. For $\left|b\right|>1$ there is no collision [\figref{pic:coll_geometry}(b)]. Finally, assuming that the particles before the encounter move with equal and constant speed, the relative angle [\figref{pic:coll_geometry}(a)]
\begin{equation} \label{eq:def_theta12}
\theta_{12} = \angle (\vec{v}_1, \vec{v}_2)
\end{equation}
suffices in place of the relative velocity. Note that $-\theta_{12}$ gives an equivalent collision geometry to $\theta_{12}$ differing only by an exchange of the particle indices.

In summary, for identical particles moving with equal speed the configuration at the moment of contact, the collision geometry, is completely determined by the impact parameter $b$ and the relative angle $\theta_{12}$.
For the scattering studies we therefore prepare the two particles in an initial state with $b \in [-1,1]$ and $\theta_{12} \in [0,\pi]$.

\subsection{Parallel alignment parameters}\label{sect:alignment_strength}

To characterize the strength of parallel alignment in binary collisions we choose the particles' mean polarization, which is hereafter referred to as  \emph{parallel alignment} 
\begin{equation}
 	A := \frac{1}{2} \left| \frac{\vec{v}_1}{\left| \vec{v}_1 \right|} + \frac{\vec{v}_2}{\left| \vec{v}_2 \right|} \right| ,
\label{def:align}
\end{equation} 
with $A \in  [0,1]$. For two particles moving exactly in the same direction $A=1$, whereas for particles moving in opposite directions $A = 0$. 
The initial value of $A$ is fully determined  by the relative angle $\theta_{12}$ between the particles' velocities and given by 
\begin{equation}
	A_\text{in} = \left| \cos \left( {\theta_{12}}/{2} \right) \right|.
\end{equation}
In order to determine the change in parallel alignment, we define the \emph{relative alignment} 
\begin{equation} \label{eq:def_relative_alignment}
 	\Delta A = A^\prime - A_\text{in} ,
\end{equation}
where $A^\prime$ denotes the parallel alignment after the collision.
$\Delta A$ depends on the collision geometry $(b, \theta_{12})$ as well as the model parameters.

Obtaining a measure for the overall alignment tendency, without referring to specific values of $\theta_{12}$ and $b$, requires integration over all possible collision geometries.
As collision events for different collision geometries are in general not equally likely to occur in a given time interval, an appropriate integration weight is required when evaluating the relative alignment for each collision geometry ($b, \theta_{12}$). 
A two-dimensional system consisting of ballistically moving uncorrelated particles in an isotropic and homogeneous state, implies that impact parameters $b$ are equally likely, and relative angles are distributed according to $|\sin(\theta_{12}/2)|$~\cite{Bertin_short, Bertin_long}.
This leads to the following definition of the \emph{alignment integral}, which characterizes the overall alignment tendency:
 \begin{equation} \label{eq:alignment_integral}
 		\langle \Delta  A \rangle = \frac{1}{4} \! \int_{-1}^{+1} \! \text{d}b \int_{0}^{\pi} \text{d}\theta_{12} \ \Delta A(b,\theta_{12}) \ \left|\sin \left( \frac{\theta_{12}}{2} \right)\right| .
\end{equation}
The systematic derivation of the weight function in terms of the unit vector $\vec{\hat{e}}$ [\eeqref{eq:def_e}] and the relative velocity $\vec v_{12}$ can be found in Appendix~\ref{appendix:alignment_integral}.  
 The \textit{alignment integral} $\langle \Delta  A \rangle$  has the following properties:
\begin{itemize}
	 \item If every collision geometry $(b, \theta_{12})$ results in parallel alignment: $A^\prime(b,\theta_{12}) = 1$ $\to$  $\langle \Delta  A \rangle = +0.5$.
 	 \item If every collision geometry $(b, \theta_{12})$ results in anti-parallel alignment: $A^\prime(b,\theta_{12}) = 0$ $\to$  $\langle \Delta  A \rangle = -0.5$.
\end{itemize}
Neither case is likely to be encountered in actual active systems since, in general, the dynamics of collisions lead to a variety of post-collisional relative angles deviating from the extremal cases of zero or $\pi$.
For fully elastic collisions of particles with initially equal speeds, the  average relative alignment has a negative value,  $\langle \Delta  A \rangle|_E = -0.05233$, as obtained by integration of the elastic collision rule (cf.~\eeqref{eq:inelastic_collrule}, \cite{Brilliantov_book}). Within the framework of binary collisions, $\langle \Delta  A \rangle > \langle \Delta  A \rangle|_E$ may be taken as a heuristic criterion constituting a prerequisite for the emergence of collective motion. 
In the following, we use $\langle \Delta A \rangle$ to analyse the ordering capabilities for each model as a function of the model's control parameters.

\subsection{Results of Binary Scattering Study}\label{sect:scattering_study}

\begin{figure*}[tb]
\begin{center}
\includegraphics[width=0.49\linewidth]{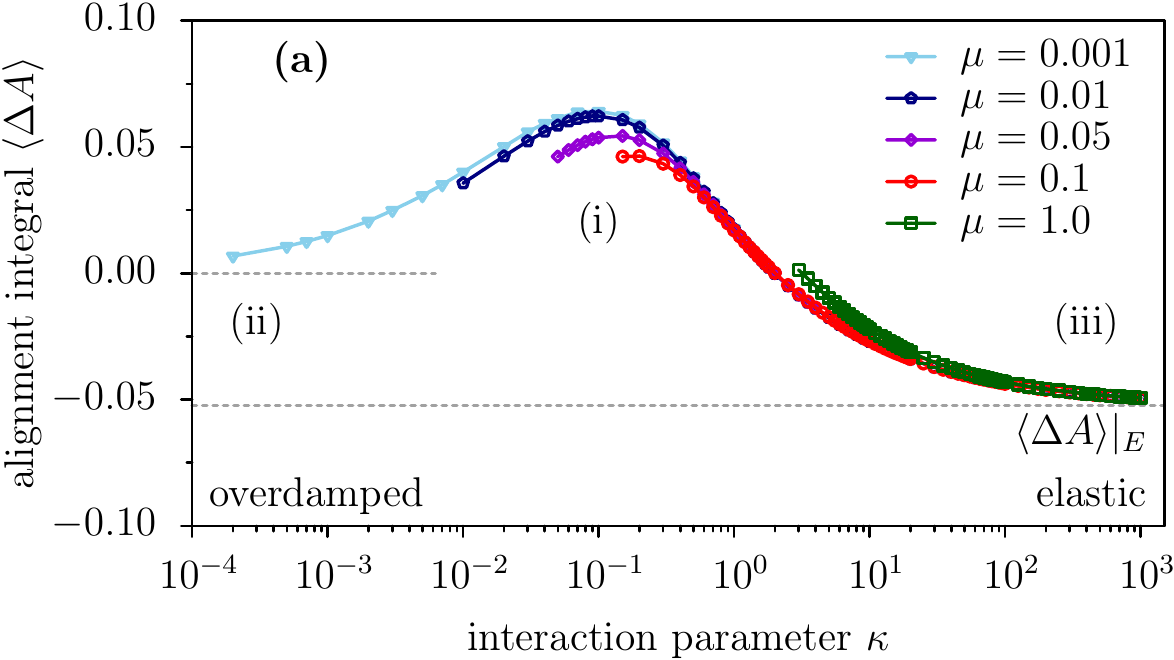}
\hfill
\includegraphics[width=0.49\linewidth]{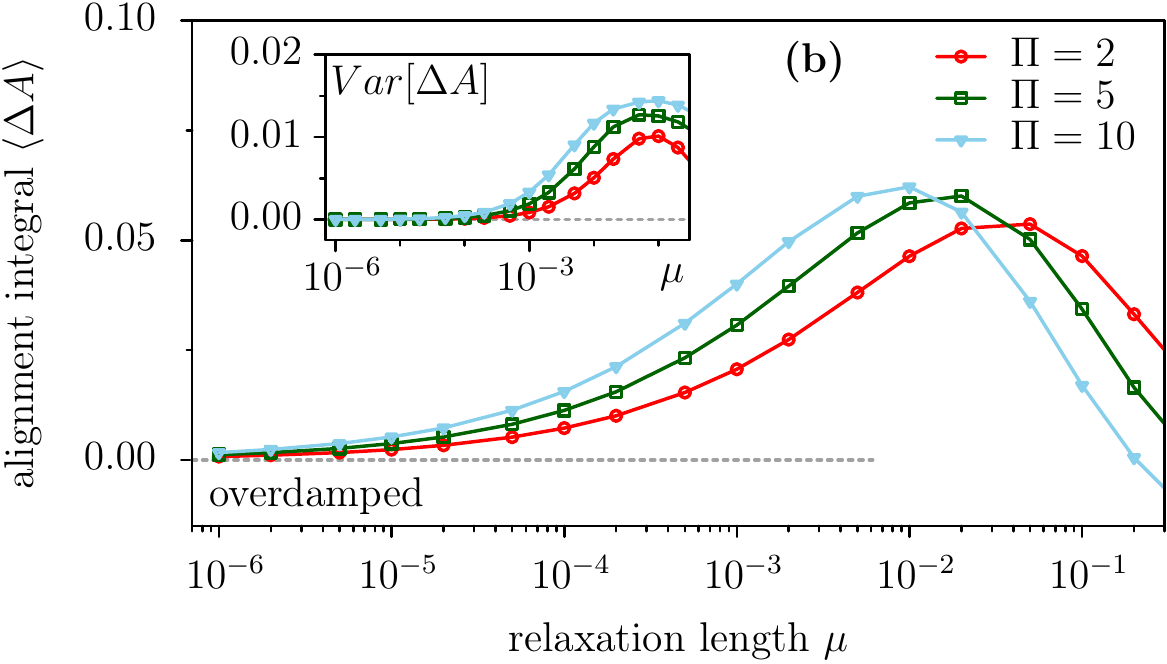}
\caption{(color online)
\textbf{(a)} $\langle \Delta A \rangle$ plotted as a function of the interaction parameter $\kappa$ for different values of the relaxation length $\mu$. Three regimes can be distinguished: (i) a pronounced maximum for intermediate values of $\kappa$ and the drop to the respective limiting values, (ii) $\langle \Delta A \rangle \to 0$ for $\kappa \to 0$ and (iii) $\langle \Delta A \rangle \to \langle \Delta A \rangle|_E=-0.05233$ in the elastic limit $\kappa \to \infty$. For each $\mu$, curves terminate for small $\kappa$, when 
the penetration distance $\xi_{ij}^\text{max}>0.9\,d$.
\textbf{(b)} $\langle \Delta A \rangle$ plotted as a function of $\mu$ for different values of the dimensionless parameter $\Phi = \kappa / \mu$ to illustrate the approach to the overdamped limit $\mu \to 0$. The inset depicts the variance $Var\left[\Delta A\right] = \langle {\Delta A}^2 \rangle - {\langle \Delta A \rangle}^2$ approaching zero for $\mu \to 0$, analog to the behavior of $\langle \Delta A \rangle$.}
\label{plot:ModelA_alignint}
\end{center}
\end{figure*}

In this section we present the results of a binary scattering study conducted by numerically integrating the respective model equations \eeqref{eq:rescaled_model_A} or \eeqref{eq:rescaled_model_B} for two particles and varying collision geometry ($b,\theta_{12}$) \CH{\footnote{\CH{For the binary scattering study, the time discretization was varied according to the demands of the respective model parameters, \emph{e.g.}\ the time step was decreased for small collision durations.}}}.

\subsubsection{Average alignment for model A}

To characterize the average alignment behavior for \emph{model A} we use the alignment integral $\langle \Delta A \rangle$ defined in \eeqref{eq:alignment_integral}. 
Varying the interaction parameter $\kappa$ for fixed relaxation length $\mu$, we find three distinct regimes, which are depicted in \figref{plot:ModelA_alignint}(a): 
(i)~A maximum in $\langle \Delta A \rangle$ at around $\kappa \approx 10^{-1}$ which corresponds to optimal alignment.
(ii)~For $\kappa \to 0$, which is equal to the overdamped limit, $\langle \Delta A \rangle$ gradually vanishes, whereas 
(iii)~for $\kappa \to \infty$, $\langle \Delta A \rangle$ approaches the elastic value $\langle\Delta A\rangle|_E$. 
Interestingly, the relaxation length $\mu$ has only a minor effect on the alignment:
For the most part, the position of the maximum of $\langle \Delta A \rangle$ is independent of  $\mu$, and its magnitude varies only weakly with the value of $\mu$ [\figref{plot:ModelA_alignint}(a)].

The aforementioned elastic limit is expected since for large $\kappa$ the impact of driving force and dissipation is negligible for the outcome of encounters [cf.\ \eeqref{eq:def_kappa}]. 
The maximum of the alignment integral always occurs for values of the interaction parameter $\kappa\approx10^{-1}$ and this value is largely independent of the relaxation length $\mu$ [\figref{plot:ModelA_alignint}(a)]. 
As $\kappa$ is given by the square of the ratio between the relaxation and the interaction timescale [\eeqref{eq:def_kappa}], $\kappa\approx10^{-1}$ implies that there is approximately a factor of $3$ between the scales on which the driving and dissipation forces and the interaction force operate.
From this we infer that achieving optimal alignment 
during a binary collision requires driving and dissipation as well as interaction forces being of similar importance, without either taking on a largely dominant role over the course of a collision. 
Note that for fixed $\mu$ the accessible range of values for the interaction parameter $\kappa$ is restricted. Decreasing $\kappa$, the relative influence of both, the driving force as well as the initial velocity \CH{(momentum)},
grows in competition to the interaction force. This results in an increase of the maximal penetration distance $\xi_{ij}^\text{max}$ during a collision and finally leads to unphysical behavior as the two particles are completely passing through each other (in a central collision).
For this reason, the curves in \figref{plot:ModelA_alignint}(a) terminate each when the maximum overlap would exceed a threshold of $\xi_{ij}^\text{max}>0.9\,d$.

Lowering the interaction parameter $\kappa$ the alignment integral $\langle\Delta  A\rangle$ decreases to zero. 
This limit is accessible only if the relaxation length $\mu \to 0$ as well. 
Therefore, in this overdamped limit both the relaxation time [\eeqref{eq:def_kappa}] as well as the relaxation length [\eeqref{eq:def_mu}] become small in relation to the respective scales of the interaction time and particle diameter.
To examine the behavior of $\langle\Delta  A\rangle$ in the overdamped limit ($\kappa,\mu \to 0$),
we introduce a new dimensionless parameter $\Pi = \kappa / \mu$. It can be interpreted as the ratio of the maximal possible interaction force $kd$ to the (constant) driving force $\alpha$. 
Then, taking the relaxation length $\mu \to 0$ is equivalent to increasing the dissipation in the system while at the same time preserving the strength ratio of the forces determining the outcome of collisions by fixing $\Pi$.
In this limit, the alignment integral $\langle \Delta  A \rangle$ decreases smoothly to zero [\figref{plot:ModelA_alignint}(b)]. 
Moreover, the variance $Var[\Delta  A] = \langle {\Delta  A}^2 \rangle - {\langle \Delta  A \rangle}^2$ vanishes identically as well [inset in \figref{plot:ModelA_alignint}(b)].
From this we conclude that $\Delta A = 0$, \emph{i.e.}, there is no change in the relative angle, not only on average but independently for any collision geometry $(b,\theta_{12})$ in the overdamped limit $\kappa,\mu \to 0$.

\subsubsection{Comparison of average alignment for model A and B}

\begin{figure}[tb]
\begin{center}
\includegraphics[width=\linewidth]{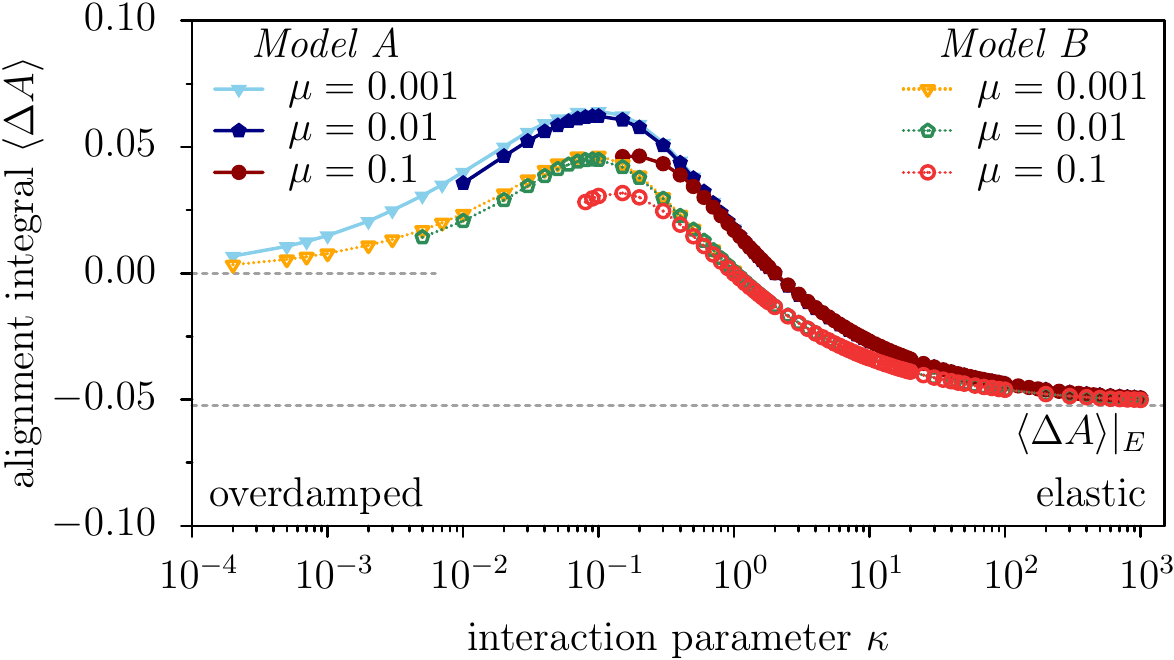}
\caption{(color online)
$\langle \Delta A \rangle$ for \emph{model A} (solid lines and filled symbols) and \emph{model B} (dotted lines and open symbols), showing the same generic shape for both models. 
 Moreover, both models have a similar behavior in the approach to the elastic limit  ($\kappa \to \infty$) as well as in the overdamped case $\langle \Delta A \rangle \to 0$  ($\kappa ,\mu \to 0$). Additionally, the position of the maximum is approximately the same for both models, however \emph{model B} exhibits generally lower values for $\langle \Delta A \rangle$.}
\label{plot:AB_comparison_alignint}
\end{center}
\end{figure}

For \emph{model A} we found that the behavior of the alignment integral $\langle\Delta A\rangle$ is comprised of three characteristic regimes. In the following we ask whether  
 \emph{model B} exhibits the same characteristics.
As \figref{plot:AB_comparison_alignint} shows, the functional form of $\langle \Delta A \rangle$ as a function of the interaction parameter $\kappa$ is indeed preserved for \emph{model B}. For $\kappa \to \infty$, the value obtained from the elastic collision rule is recovered, $\langle \Delta  A \rangle \to \langle \Delta  A \rangle|_\text{E}$. In the same way there is a convergence to a vanishing change in alignment, $\langle \Delta A \rangle \to 0$, in the overdamped case  ($\kappa, \mu \to 0$).
Furthermore, the maximum of the alignment integral $\langle \Delta A \rangle$ indicating optimal alignment occurs for values of $\kappa\approx10^{-1}$, with only little variation with the relaxation length $\mu$, identical to the behavior for \emph{model A}.
We therefore conclude that it is indeed the relative influence of the driving/dissipation forces and the interaction force that determines the aligning capabilities inherent in the model equations: 
Good alignment can only arise if neither force is particularly dominant during collisions, indicated by an intermediate value of $\kappa$ in \figref{plot:AB_comparison_alignint}.

The only difference between the models is found in the magnitude of the alignment integral $\langle \Delta A \rangle$ at the maximum [\figref{plot:AB_comparison_alignint}]:  
The amplitudes of the maxima for \emph{model B} are lower for all values of $\mu$.
Understanding the reason for this difference requires investigation of the change in alignment $\Delta A(b,\theta_{12})$ as a function of the collision geometry $(b,\theta_{12})$. This analysis is detailed in the next section and will also reveal the underlying mechanism responsible for the maximum in $\langle \Delta A \rangle$.

\protect{\subsubsection{Alignment for model A and model B: \\ the collision rule}}

\begin{figure*}[tbp]
\begin{center}

\includegraphics[width=\linewidth]{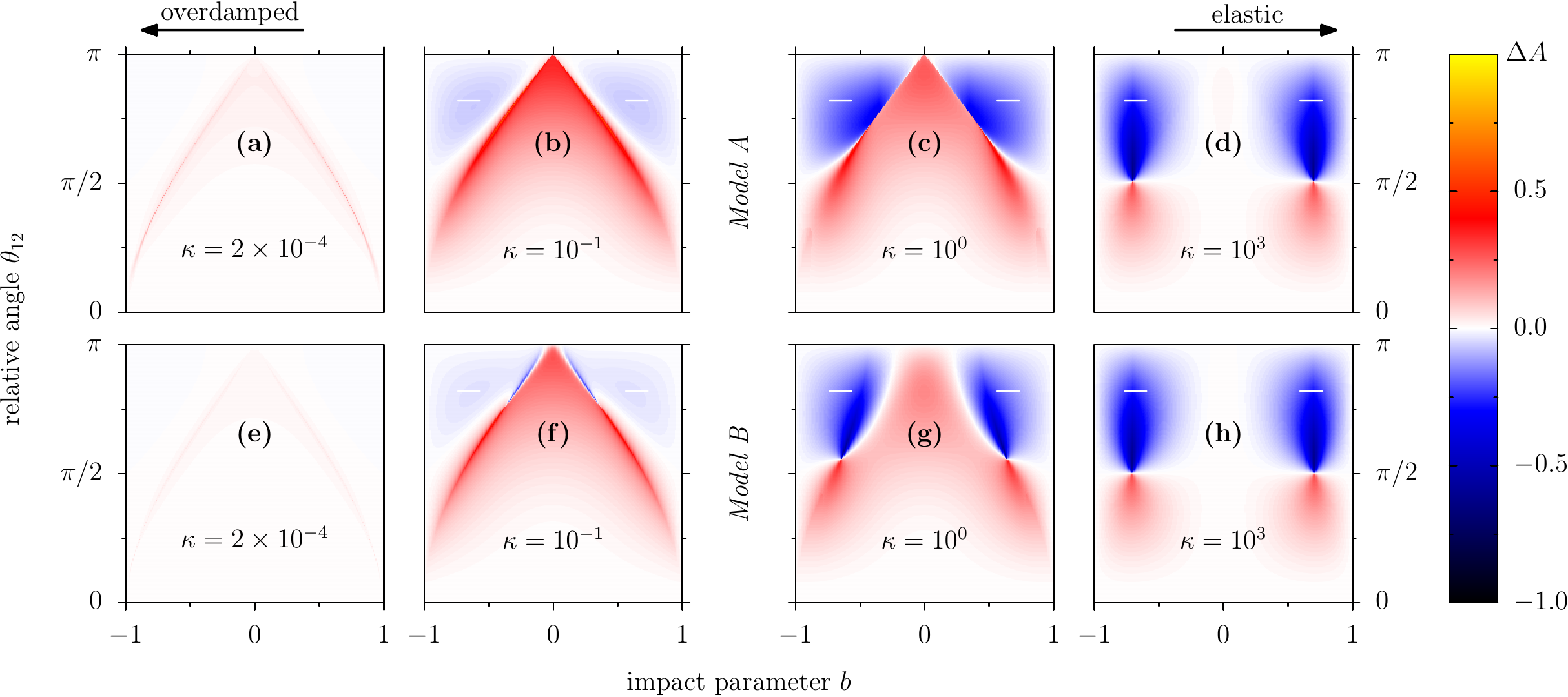}
\caption{
(color online) Change in alignment $\Delta A$ as a function of the collision geometry given by the relative angle $\theta_{12}$ and the impact parameter $b$ for \emph{model A} (top row, \textbf{a}--\textbf{d}) and \emph{model B} (bottom row, \textbf{e}--\textbf{h}) with $\mu = 0.001$ and varying $\kappa$ as denoted in the figure. 
Blue/black (dark grey) regions (marked with a minus sign) denote collisions with post-collisional alignment $A^\prime < A$, while for red/yellow (unmarked, lighter grey) regions $A^\prime > A$. White signals no change as denoted by the color-code on the right. 
\textbf{(a,e)} For both models, very low values of $\kappa$ show a distinct "washed-out" quality, coinciding with the approach to the overdamped limit, where positive as well as negative change disappears for all collision geometries.
\textbf{(d,h)} For $\kappa = 10^{3}$, behavior is very close to fully elastic collisions showing very pronounced regions of $\Delta A < 0$ for $\theta_{12} > \pi /2$ (marked with a minus sign). 
\textbf{(c,g)} Increasing $\kappa$ leads to the regions of positive alignment spreading to larger $\theta_{12}$, while those of negative alignment recede, until around the maximum of $\langle \Delta A \rangle$ ($\kappa = 10^{-1}$),
\textbf{(b,f)} there are pronounced peaks of positive $\Delta A$ for all $\theta_{12}$, forming the edges of an almost triangular shape. 
}
\label{plot:ModelAB_alignmaps}
\end{center}
\end{figure*}

So far we have studied the binary scattering behavior on average by means of the alignment integral $\langle \Delta A \rangle$, which represents a mean value over all collision geometries $(b,\theta_{12})$. 
Now, we take a closer look and study the change in alignment $\Delta A(b,\theta_{12})$  as a function of the specific collision geometry given by the relative angle $\theta_{12}$, and the impact parameter $b$.
Choosing a fixed value for the relaxation length $\mu=0.001$, 
\figref{plot:ModelAB_alignmaps} depicts $\Delta A(b,\theta_{12})$ for \emph{model A} (a-d) and \emph{model B} (e-h)
for values of the interaction parameter $\kappa \in \{2\times10^{-4}, 10^{-1}, 10^{0}, 10^{3}\}$, respectively. 
The largest value of $\kappa$ corresponds to mostly elastic collisions, and the smallest value is close to the overdamped limit. Recall that the maximum in the alignment integral $\langle \Delta A \rangle$ [\figref{plot:AB_comparison_alignint}] is reached for $\kappa=10^{-1}$.

\begin{figure}[tbp]
\begin{center}
\includegraphics[width=\linewidth]{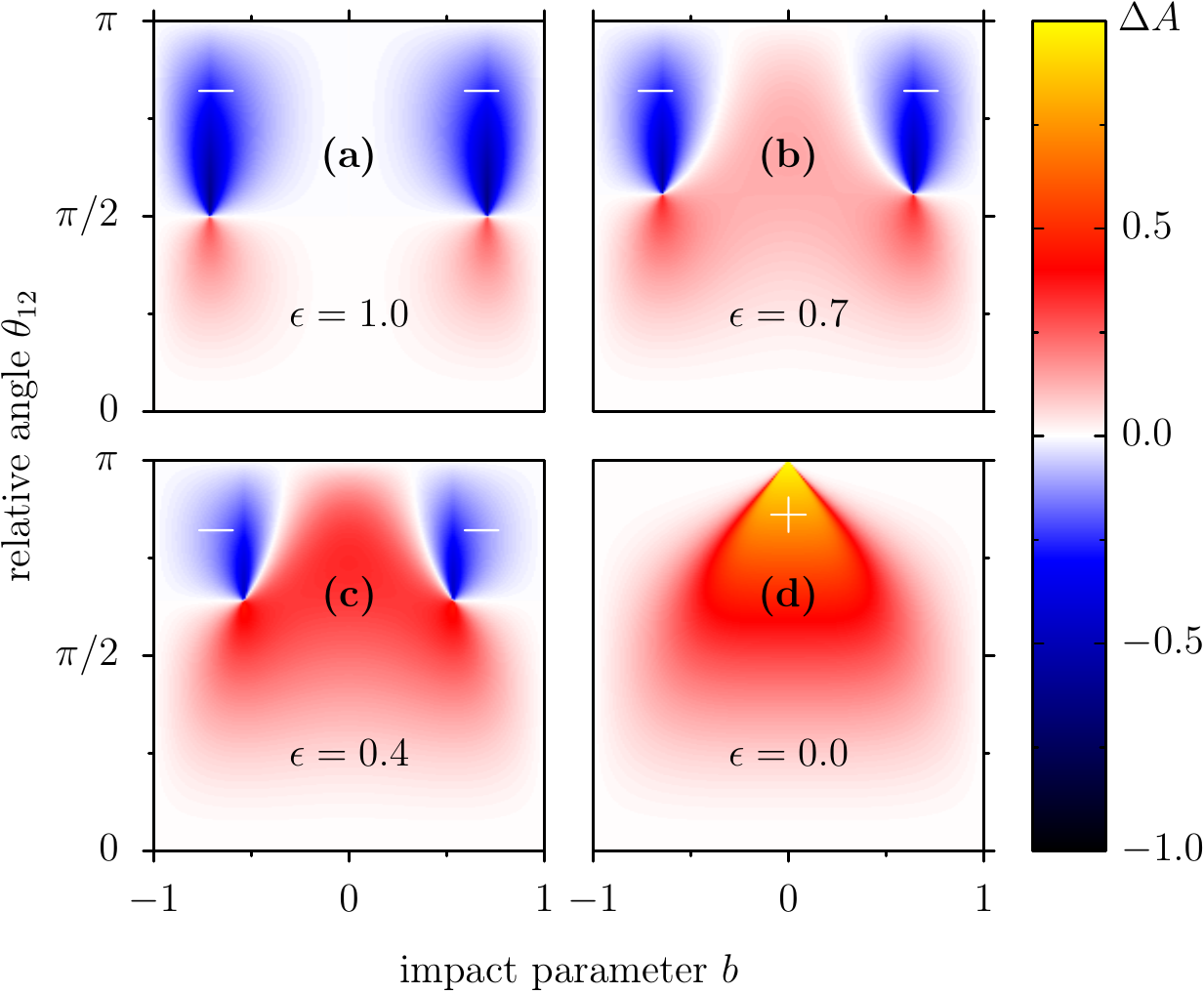}
\caption{(color online) Change in alignment $\Delta A$ as a function of the collision geometry given by the relative angle $\theta_{12}$ and the impact parameter $b$ obtained by use of the inelastic collision rule [\eeqref{eq:inelastic_collrule}]. The normal coefficient of restitution $\epsilon$ varies between a fully elastic collision [$\epsilon=1$, \textbf{(a)}] and a fully inelastic collision [$\epsilon=0$, \textbf{(d)}]. Blue/black (dark grey) regions (marked with a minus sign) denote collisions with post-collisional alignment $A' < A$, while for red/yellow (unmarked, lighter grey) regions $A' > A$ (see mapping of $\Delta A=A'-A$ to corresponding shading on the right). For a video please refer to the Supplementary Material~\cite{Supplement_Videos}.  
}
\label{plot:collrule_alignmaps}
\end{center}
\end{figure}

The change in alignment $\Delta A(b,\theta_{12})$ for \emph{model A}  and \emph{model B} features strong similarities 
(see also a video in the Supplemental Material~\cite{Supplement_Videos}):
For a large interaction parameter ($\kappa = 10^3$), $\Delta A(b,\theta_{12})$ exhibits balloon-like regions with rather large values, located at intermediate values of the impact parameter $b \approx \pm \sqrt{2}/2$. 
At a relative angle $\theta_{12}\approx\frac{\pi}{2}$, corresponding to orthogonal pre-collisional velocities, there is a discontinuous change from positive $\Delta A$ for smaller $\theta_{12}$ to negative values for larger $\theta_{12}$ [Figs.~\ref{plot:ModelAB_alignmaps}(d) and \ref{plot:ModelAB_alignmaps}(h)]. 
This is the result of a flip in the post-collisional velocity of one of the colliding particles
(see Video in Supplemental Material~\cite{Supplement_Videos}, 
\footnote{For $\theta_{12}<\frac{\pi}{2}$ and $b\approx\pm\frac{\sqrt{2}}{2}$, particles' velocities are aligned in elastic collisions, with one particle's velocity being increased and the other's becoming small. The increase in the relative velocity $\left|\vec{v}_{12}\right|$ when going to larger relative angles $\theta_{12}>\frac{\pi}{2}$ results in the latter particle's velocity being completely flipped. The former particle's scattering behavior changes in a continuous manner, resulting in the discontinuous change in $\Delta A$ observed in Figs.~\ref{plot:ModelAB_alignmaps}(d) and \ref{plot:ModelAB_alignmaps}(h).}). 
As discussed further below these results for $\kappa = 10^3$ are almost identical to those found for perfectly elastic collisions; see \figref{plot:collrule_alignmaps}(a). 
For small values of $\kappa= 2\times10^{-4}$ [Figs.~\ref{plot:ModelAB_alignmaps}(a) and \ref{plot:ModelAB_alignmaps}(e)], all features present at larger values of $\kappa$ essentially vanish, and  $\Delta A(b,\theta_{12}) \approx 0$ for all collision geometries. 
 This result has already been implied by the analysis of the alignment integral $\langle\Delta A\rangle$ and its variance, which both approach zero in the overdamped limit [\figref{plot:ModelA_alignint}].

For intermediate values of the interaction parameter $\kappa$, there is a triangular-shaped region of positive $\Delta A$ across the whole range of relative angles $\theta_{12}\in[0,\pi]$ [see Figs.~\ref{plot:ModelAB_alignmaps}(b) and~\ref{plot:ModelAB_alignmaps}(f)  for $\kappa=10^{-1}$, and to a lesser extent Figs.~\ref{plot:ModelAB_alignmaps}(c) and \ref{plot:ModelAB_alignmaps}(g) for $\kappa=10^{0}$].
At two edges of this triangular region, there are pronounced peaks of positive $\Delta A$, which for $\kappa=10^{-1}$ are the most prominent features in the graph, and therefore provide the dominant contribution to the alignment integral [\figref{plot:AB_comparison_alignint}].

\begin{figure}[tbp]
\begin{center}
\begin{tikzpicture}
    \node[anchor=south west,inner sep=0] (image) at (0,0) {\includegraphics[width=0.75\linewidth]{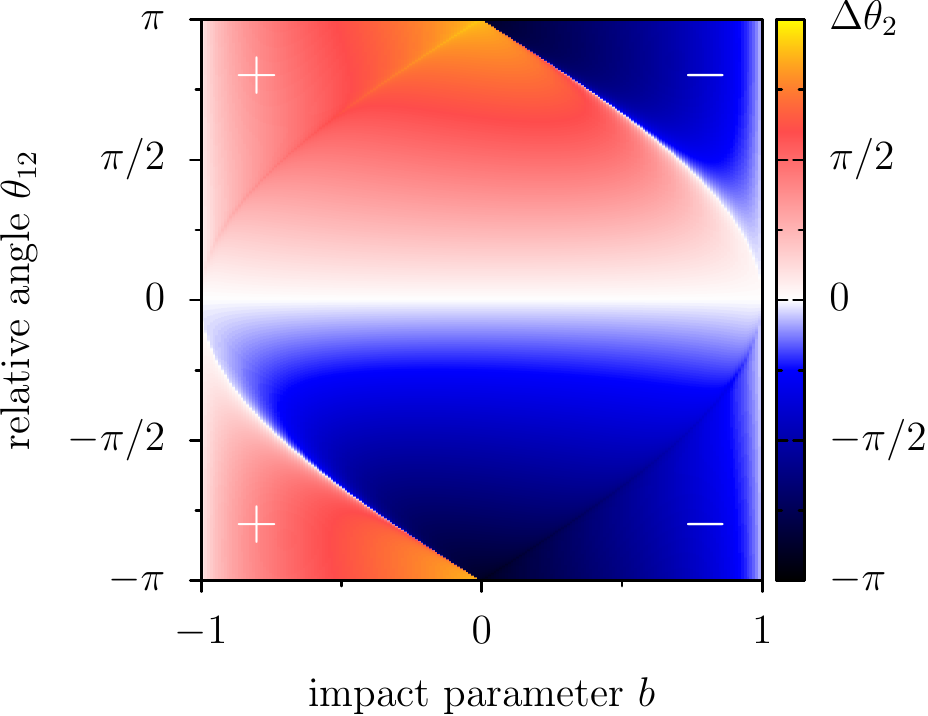}};
    \begin{scope}[x={(image.south east)},y={(image.north west)}]
        \node at (0.07,0.97) {\textbf{(a)}};
        \node at (1.05,0.97) {\textbf{(b)}};
        \node[coordinate] (sketch) at (1.18,0.7) {};
    \end{scope}
    \begin{scope}[scale=0.35, rotate=-90, dashedlines/.style={thick,color=gray,densely dashed}, lines/.style={thick,color=black}, circles/.style={thick,color=black},]
	\draw[circles] (sketch) ++(0,0) node[coordinate] (zero) {} +(180:2.5) circle (1) node[coordinate] (two) {} +(100:1.9) circle(1) node[coordinate] (one) {};
	\draw[dashedlines] (zero) -- (two) (zero) -- (one);
	\draw[lines,->] (two) -- +(0:1.5) node[left] {$\vec{v}_{2}$};
	\draw[lines,->] (one) -- +(-80:1.5) node[below] {$\vec{v}_{1}$};
	\draw[very thick,color=black,double,>=latex,->] (zero) ++(1.5,0.5) -- +(2,0);
	\draw[circles] (zero) ++(5,0) node[coordinate] (zero) {} +(180:0) circle (1) node[coordinate] (two) {} +(35:2) circle(1) node[coordinate] (one) {};
	\draw[lines,->] (two) -- +(-5:1.5) node[left] {$\vec{v}^\prime_{2}$};
	\draw[lines,->] (one) -- +(-5:1.5) node[right] {$\vec{v}^\prime_{1}$};
    \end{scope}
\end{tikzpicture}
\caption{(color online)
\textbf{(a)} Change in the direction of motion of particle $2$,
 $\Delta\theta_2$, called scattering angle,
 as a function of the collision geometry $(\theta_{12}, b)$.
Parameters: \emph{model A}, $\mu=0.001$, $\kappa=10^{-1}$, corresponding to a maximum in $\langle \Delta A \rangle$.
As both particles are identical spheres, the scattering behavior of the particle $1$ can be read off at the point $\left({-\theta_{12}},b\right)$ in the plot. 
\textbf{(b)} Schematic illustration of the dominant alignment principle termed as ``\emph{alignment of the first}". During the collision $\vec{v}_1$ is aligned to $\vec{v}_2$ with only marginal change of $\vec{v}_2$. }
\label{plot:oneparticlescattering_map}
\end{center}
\end{figure}

Understanding the detailed scattering behavior for parameters corresponding to the edges of the triangular-shaped region is hence vital for determining the underlying principle of alignment.
Let $\theta_i$ be the angle of the pre-collisional velocity for particle $i$ with respect to some reference axis, and $\theta_i^\prime$ the corresponding angle after the collision. Then, the scattering angle $\Delta \theta_i=\theta_i^\prime-\theta_i$ describes the change in the particle's direction of motion as the result of a collision with another particle. 
Figure~\ref{plot:oneparticlescattering_map}(a) shows the scattering angle $\Delta\theta_2$ for particle $2$ as a function of the collision geometry ($b, \theta_{12}$) for \emph{model A} and parameters $\mu=0.001$, $\kappa=10^{-1}$.
The scattering angle $\Delta\theta_1$ for particle $1$ in the same collision can be read off at the point ($b,-\theta_{12}$) in \figref{plot:oneparticlescattering_map}(a); this can be seen by considering an exchange of indices in the definitions of the relative angle $\theta_{12}$ [\eeqref{eq:def_theta12}] and the impact parameter $b$ [\eeqref{eq:def_impact}].
The scattering angle $\Delta \theta_2$ in \figref{plot:oneparticlescattering_map}(a) exhibits the same kind of triangular structure as found for $\Delta A$ in \figref{plot:ModelAB_alignmaps}(c).
For collision geometries at the edges of this triangular structure, one particle hardly changes its direction of motion [white region in \figref{plot:oneparticlescattering_map}(a)], while the orientation of the other particle changes by an angle close to the relative angle $\theta_{12}$.
This results in an alignment of the latter particle's velocity to that of its collision partner (see Supplementary Material~\cite{Supplement_Videos} for a video).
Closer examination of the collision geometry reveals that it is the ``first'' particle's velocity that is aligned. Here, ``first'' means that before the collision it is closer to the \emph{center of collision} [see \figref{pic:coll_geometry}(a)], defined by the intersection point of the  pre-collisional orientations [particle $1$ for $b>0$, see the sketch in \figref{plot:oneparticlescattering_map}(b)]. 
The orientation of the other particle [particle $2$ in \figref{plot:oneparticlescattering_map}(b)] hardly changes 
 because the repulsive force mostly affects 
 the magnitude of its velocity, which is counteracted to some extent by the driving force. At the same time, the ``first'' particle's velocity is rotated quickly until both velocities become aligned.
Therefore, we term this mechanism ``\emph{alignment of the first}.'' 
Interestingly, all collision geometries corresponding to the edges of the triangular structure lead to ``\emph{alignment of the first}'' (refer to the Supplemental Material~\cite{Supplement_Videos} for a video).
Following the same line of reasoning for \emph{model B}, we find the identical principle of alignment. We conclude that the ``\emph{alignment of the first}'' is the dominant mechanism giving rise to parallel alignment during binary collisions of soft active colloids.

Finally, we contrast the scattering behavior of particles governed by the equations of motion of \emph{model A} or \emph{model B} (for all values of $\kappa$) to that of classical inelastic collisions described by the inelastic collision rule with a fixed normal coefficient of restitution $\epsilon$.
Denoting the velocities of two particles after a collision as $\vec{v}_1^\prime$ and $\vec{v}_2^\prime$,  the classical 
 inelastic collision rule is given by \cite{Brilliantov_book}:
\begin{equation} \label{eq:inelastic_collrule}
 \begin{split}
  \vec{v}_1^\prime & = \vec{v}_1 - \frac{1+\epsilon}{2} \left( \vec{v}_{12} \cdot \vec{\hat{e}} \right) \vec{\hat{e}} ,\\
  \vec{v}_2^\prime & = \vec{v}_2 + \frac{1+\epsilon}{2} \left( \vec{v}_{12} \cdot \vec{\hat{e}} \right) \vec{\hat{e}} .
 \end{split}
\end{equation}
The normal coefficient of restitution $\epsilon$ determines the change in the normal component of the relative velocity $\vec{v}_{12}$: $\epsilon=-(\vec{v}_{12}^\prime\cdot\vec{\hat{e}})/(\vec{v}_{12}\cdot\vec{\hat{e}})$.
The relative alignment $\Delta A$ for a collision with collision geometry ($b, \theta_{12}$) is shown in \figref{plot:collrule_alignmaps} for values of $\epsilon$ ranging from fully elastic collisions ($\epsilon=1$), to collision where the normal component of the relative velocity is damped completely ($\epsilon=0$).
Similar to \emph{model A} and \emph{model B} [Figs.~\ref{plot:ModelAB_alignmaps}(b) and~\ref{plot:ModelAB_alignmaps}(f)] we find for very small $\epsilon$ a triangle-like region of positive $\Delta A$ [\figref{plot:collrule_alignmaps}(d)], however the edges of the triangle are far less pronounced.
For inelastic collisions, most of the contribution to the alignment integral $\langle\Delta A\rangle$ comes from collision geometries corresponding to the inner region of the triangle, \emph{i.e.}, small $|b|$.
 For these small impact parameters the relative velocity is directed along the normal direction and vanishes in a fully inelastic collision.
 In contrast, \emph{model A} and \emph{model B} exhibit only small $\Delta A$ for small impact parameters  [Figs.~\ref{plot:ModelAB_alignmaps}(a-d) and \ref{plot:ModelAB_alignmaps}(e-h)].

Comparing inelastic collisions with \emph{model A} at fixed $\langle\Delta A\rangle$ corresponding to \emph{model A}'s maximum, it turns out that inelastic collisions still exhibit the balloon-like structure in $\Delta A(b,\theta_{12})$, while \emph{model A} already developed the triangle (see video in the Supplemental Material~\cite{Supplement_Videos}).
Taken together, we conclude that
\emph{model A} (and \emph{model B}) follow an alignment mechanism that is \CH{distinctively different from that governing simple inelastic collisions described by a constant restitution coefficient}: 
Alignment in the inelastic collision rule comes from damping of the normal component of the relative velocity, whereas for the dynamic models it is the presence of the propelling force that keeps one particle ``stuck'' against the repulsive force and enables the alignment of the other particle's velocity according to the ``\emph{alignment of the first}" mechanism.

In summary, the scattering study provided the means to systematically study the alignment properties of soft active colloids.
\CH{In these systems the interplay of driving, dissipative and repulsive interaction forces introduces nonlinearities in the dynamics, giving rise to alignment between colliding particles.}
We found that the nature of collisions is determined by two dimensionless parameters: the interaction parameter $\kappa$, which determines the influence of the driving force and dissipation relative to the interaction force during collisions, and the relaxation length $\mu$, which gives the typical relaxation length relative to the particle size.
We observed that each model comprises two distinct limits determined by these dimensionless parameters: the \emph{elastic limit} ($\kappa\to\infty$), where particles obey an elastic collision rule, and the \emph{overdamped limit} ($\kappa,\mu\to0$), where there is no change to relative orientations. 
Further, we identified the model parameters for which parallel alignment is maximal, 
and found that this maximum occurs for all models at the same intermediate value of the interaction parameter $\kappa$, largely independent of $\mu$.
Additionally, parallel alignment for all considered models followed the same principle, termed  ``\emph{alignment of the first}", regardless of the type of driving or dissipating force. 
The principle states that those collision geometries that contribute dominantly to the particles' parallel alignment exhibit the following typical characteristics in the particles' dynamics: 
the first incoming particle [with respect to the \emph{center of collision},  \figref{pic:coll_geometry}(a)] is aligned parallel to the second of the colliding particles.
\CH{Moreover, we showed that this alignment principle for soft active colloids is distinctively different from simple inelastic granular gases described by a constant restitution coefficient.} 
Identification of the universal fingerprints of colliding active colloids, as well as the conditions for maximal alignment, 
provides the starting point for a study of the collective properties of active colloids. On the basis of the ``collision-rule'' depicted in \figref{plot:oneparticlescattering_map}(a), we derive a mesoscopic description using the framework of kinetic theory for propelled particle systems.

\section{Scrutinizing kinetic theory for propelled particles}\label{sect:collective}

In order to connect  the system's collective behavior studied in section~\ref{sect:multi_particle_simulation}  with the results of the binary scattering study, we use kinetic theory for propelled particles moving at constant speed~\cite{Bertin_short, Bertin_long}. Kinetic theory aims to provide a description for the time evolution of the one-particle distribution function $f(\vec r, \theta,t)$, which is a function of the spatial coordinates $\vec r$, the orientation of the velocity $\theta$ and time $t$. It is conceptually restricted to binary interactions between the constituent particles, limiting its range of validity to dilute conditions (packing fraction $\Phi\ll1$).  The binary interactions are described by collision integrals with each kernel involving a measure for the rate of collisions, known as \emph{Boltzmann collision cylinder}, as well as a ``collision rule.'' The latter constitutes a mapping between the pre-collisional 
angles $\theta_1$ and $\theta_2$  
and the post-collisional orientations of the two colliding particles. The corresponding distribution function required to compute the rate of binary collisions is the two-particle density $f^{(2)}(\vec r, \theta_1,\theta_2,t)$. To obtain a closed equation for the time evolution of the one-particle density $f(\vec r, \theta,t)$, called the \emph{Boltzmann equation}, the \emph{assumption of molecular chaos} is commonly made~\cite{Bertin_short, Bertin_long}, \emph{i.e.}, one assumes that correlations in space and orientation are completely absent such that 
\begin{equation}\label{eq:mol_chaos}
 		f^{(2)}(\vec r, \theta_1,\theta_2,t)=f(\vec{r},\theta_1,t) f(\vec{r},\theta_2,t).
\end{equation} 
This constitutes a rather strong assumption concerning the system's dynamics and it is not clear to which degree this assumption holds for active systems, in particular at the onset of collective motion.

For the following analysis we restrict ourselves to \emph{model~A}  because, as we have shown in the last section, \emph{model~A} and \emph{model~B} are equivalent with respect to their qualitative alignment principle. 
Moreover, we specify a certain parameter set, namely $(\mu, \kappa)=(0.05, 0.1)$ \footnote{The value of $\kappa$ was chosen to be close to the maximum in {$\langle \Delta A \rangle$}. The value of $\mu$ was chosen out of consideration for computational efficiency.}, which corresponds to the maximum of the alignment parameter $\langle \Delta A \rangle$
[see \figref{plot:ModelA_alignint}(a)], and is equal to the parameter set used for the multi-particle simulations described in section~\ref{sect:multi_particle_simulation}.
We expect that optimal alignment in a binary collision also optimizes the capability of a multi-particle system to develop a macroscopic polarized state. 
Moreover, if polar order develops the critical packing fraction should be lowest for optimal alignment. This in turn improves the validity of a Boltzmann description as the regimes of large packing fractions are expected to be captured insufficiently by this kinetic approach.

\subsection{Coarse grained collision rule}\label{sect:collective_coarse_grained_col_rule}

\begin{figure}[tbrl]
\begin{center}
\includegraphics[width=\linewidth]{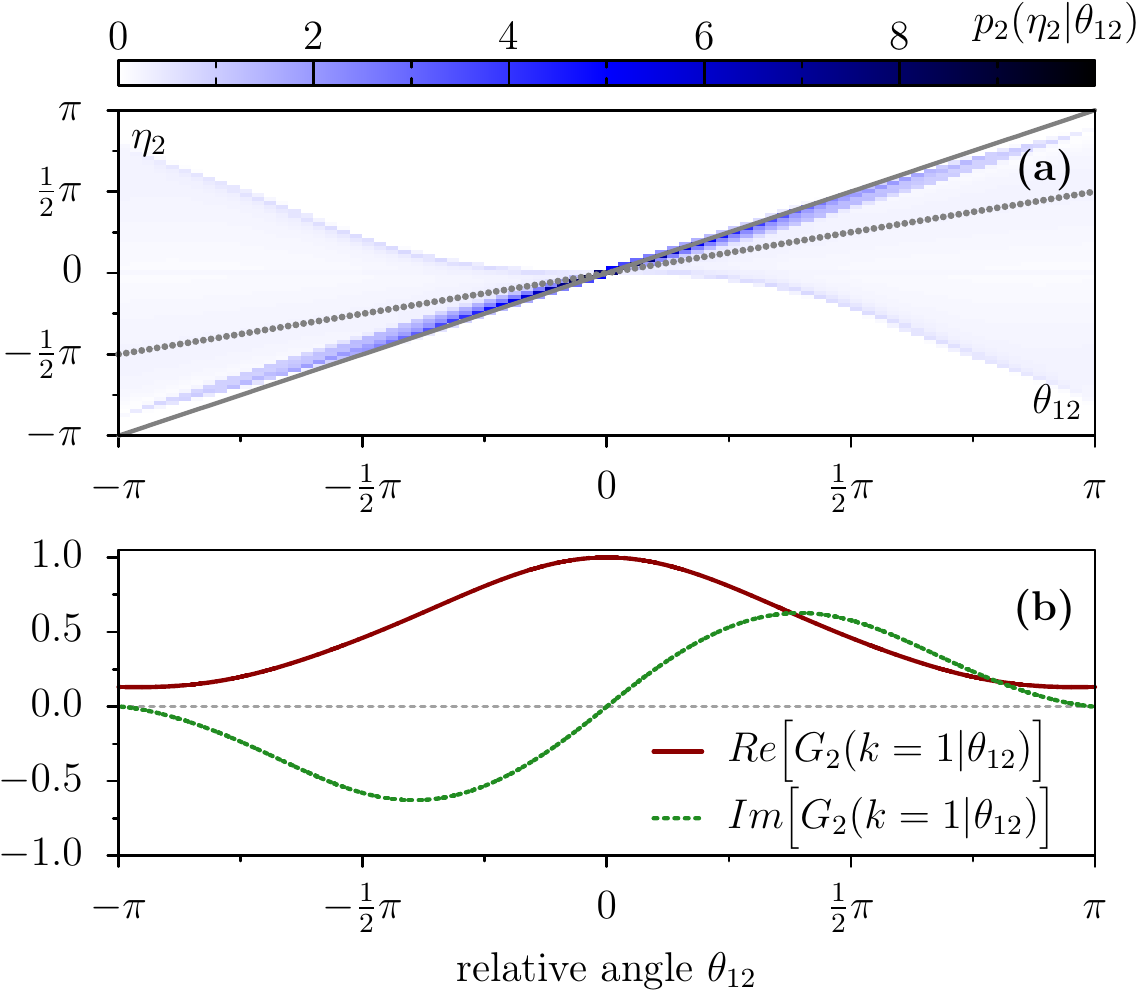}
\caption{(color online)
\textbf{(a)} \emph{Distribution} $p_2(\eta_2|\theta_{12})$ for the scattering angle $\eta_2$ of particle $2$ as a function of the relative angle $\theta_{12}$ obtained by a reduction of variables with respect to the impact parameter $b$ [see \eeqref{eq:scatDist}]. The maximum of the distribution deviates slightly from the case of specular reflection where particles exchange their orientations (solid grey line). The dotted line marks the scattering angle of a half-angle alignment rule \cite{Aronson_MT, Bertin_short, Bertin_long}.
\textbf{(b)} \emph{Real and imaginary parts of the characteristic function} $G_2(k|\theta_{12})$ of the  distribution $p_2(\eta_2|\theta_{12})$ for $k=1$ as a function of the relative angle $\theta_{12}$.
Parameters:\ \emph{Model~A} with $\mu=0.05$ and $\kappa=10^{-1}$. 
}
\label{plot:scatDist_charFunct}
\end{center}
\end{figure}

The collision rule required to set up the Boltzmann equation maps the pre-collisional orientations given by the angles $\theta_1$ and $\theta_2$ on the post-collisional orientations, denoted as $\theta_1^\prime$ and $\theta_2^\prime$.
Denoting the angular change for particle  $j\in\{1,2\}$ by $\eta_j(\theta_{12})$,
 the collision rule has the following general form
\begin{equation} \label{eq:coll_rule_general}
	\big(\theta_1, \, \theta_2\big) \to \big(\theta_1+\eta_1(\theta_{12}), \, \theta_2+\eta_2(\theta_{12})\big). 
\end{equation}
In a collision with given relative pre-collisional angle $\theta_{12}$,
a scattering angle  $\eta_j(\theta_{12})$ occurs with probability $p_j(\eta_{j}|\theta_{12})\text{d}\eta_j$, where
 \begin{equation}\label{eq:scatDist}
	p_j(\eta_{j}|\theta_{12}) = \frac{1}{2}\int_{-1}^{+1} \text{d}b \ \delta\big(\Delta\theta_j(b,\theta_{12}) - \eta_{j} \big).
\end{equation}
Since $p_j(\eta_{j}|\theta_{12})$ is computed from $\Delta\theta_j(b,\theta_{12})$ (see \figref{plot:oneparticlescattering_map}(a), \footnote{Note that \figref{plot:oneparticlescattering_map}(a) depicts the scattering angle for a different value of $\mu$.}) by integrating over the impact parameter $b$, we have now turned from a deterministic description to a probabilistic treatment of the collision process. As shown  in \figref{plot:scatDist_charFunct}(a), $p_2(\eta_{2}|\theta_{12})$ has a pronounced maximum 
at $\eta_2(\theta_{12})= m \theta_{12}$ with $m \lesssim 1$ for most of the range of relative angles $\theta_{12}\in[-\pi,\pi]$. This maximum is close to specular reflection ($m=1$), yet is skewed towards a slightly smaller post-collisional relative angle. However, the maximum is far removed from a half-angle alignment rule where 
$\left(\theta_1, \, \theta_2\right) \to \left(\bar\theta
, \bar\theta\right)$ with $\bar \theta=(\theta_1+\theta_2)/2$~\cite{Aronson_MT, Bertin_short, Bertin_long}, leading to $\eta_2=\theta_{12}/2$ [see dotted line in \figref{plot:scatDist_charFunct}(a)].
Overall, the distribution $p_j(\eta_{j}|\theta_{12})$ indicates that our model exhibits a  bias towards alignment, even though it is rather weak.

Note that assigning indices $1$ and $2$ to the individual particles in a collision is a matter of convention yet affects the sign of $\theta_{12}$. Exchange symmetry between the identical particles then enforces that
$p_1(\eta_1|\theta_{12})=p_2(\eta_2|{-\theta_{12}})$.
Additionally, consider collisions with relative angles $\theta_{12}$ and $-\theta_{12}$, respectively, each seen from the point of view of a specific particle, say particle~$2$. Since
these two pre-collisional states exhibit a mirror symmetry with respect to the velocity direction of particle $2$, the outcome of the collisions for particle $2$ differs only by the sign of its scattering angle, \emph{i.e.}, $p_2(\eta_2|{-\theta_{12}})=p_2(-\eta_2|\theta_{12})$. The same argument applies for particle $1$. Taken together, we have:
\begin{equation}\label{eq:scatDist_symmetry}
	p_1(\eta_{1}|\theta_{12}) = p_2(-\eta_{2}|\theta_{12}) ,
\end{equation}
\emph{i.e.}, given a collision with a relative angle $\theta_{12}$, the respective  distributions of the two particles are related by a change in the sign of the argument.

Later, in the analysis of the Boltzmann equation in Fourier space the characteristic functions of the distributions [\eeqref{eq:scatDist}, \figref{plot:scatDist_charFunct}(a)] will be required, which are defined as
\begin{equation} \label{eq:def_charfkt}
	G_j(k|\theta_{12}) = \int_{-\infty}^{\infty} \text{d}\eta_j \, e^{i k \eta_j} p_j(\eta_j|\theta_{12}) .
\end{equation}
Due to the symmetry between the distributions for the two particles [\eeqref{eq:scatDist_symmetry}] the corresponding characteristic functions are complex conjugates of each other:
\begin{equation} \label{eq:charfkt_cc}
G_1(k|\theta_{12}) = \big(G_2(k|\theta_{12})\big)^{*} ,
\end{equation}
where $^{*}$ denotes complex conjugation. The real and imaginary parts of $G_2(k|\theta_{12})$ are depicted in  \figref{plot:scatDist_charFunct}(b) for $k=1$.
Instead of using the full characteristic function for the Boltzmann equation, it is tempting to account solely for the average and the deviation computed for the  distributions $p_j(\eta_{j}|\theta_{12})$ defined in \eeqref{eq:scatDist} (\emph{Gaussian approximation}). However, such a procedure fails because it  misrepresents the actual scattering behavior by underestimating the impact of scattering events with large relative angles, and thus masks the true reason for the emergence of collective motion (see appendix \ref{appendix:gaussian_approximation} for details).

\subsection{Kinetic approach based on coarse grained collision rule}\label{sect:collective_kinetic_model}

Following Refs.\ \cite{Bertin_short, Bertin_long}, the time evolution of the one-particle distribution function $f(\vec{r},\theta,t)$ is  determined by a Boltzmann equation of the form
\begin{subequations}\label{eq:BoltzmannEquation_all}
\begin{equation}\label{eq:BoltzmannEquation}
\partial_t f(\vec{r},\theta,t)+v_0\vec{\hat{v}}(\theta)\cdot\nabla f(\vec{r},\theta,t)=\mathcal{N}[f]+\mathcal{C}[f^{(2)}] .
\end{equation}
The second term on the l.h.s.\ in \eeqref{eq:BoltzmannEquation} is the streaming term accounting for movement of particles with the velocity $v_0\vec{\hat{v}}(\theta)$, where $\vec{\hat{v}}(\theta)=\left(\cos\theta, \sin\theta\right)$.
The speed $v_0$ is assumed to be a constant
\footnote{Since the value of the dimensionless relaxation length $\mu=0.05$ is small, according to its definition in \eeqref{eq:def_mu}, particle velocities relax quickly to the stationary speed $v_0$ over a distance small compared to the particle diameter. Thus, on the coarse-grained scale of a kinetic model a constant particle speed is a good approximation.} 
and taken equal to the stationary speed $v_0=\alpha/\beta$, with $\alpha$ and $\beta$ denoting the amplitudes of the driving and friction forces in the microscopic model, respectively [\eeqref{eq:constitutive_model_A}]. The term $\mathcal{N}[f]$ describes single particle, angular fluctuations (self-diffusion events) occurring at a rate $\lambda$ and reads \cite{Bertin_short, Bertin_long}
\begin{align}
	\begin{split}
	\mathcal{N}[f] = & - \lambda f(\vec{r},\theta,t)  
	+ \lambda \int_{-\pi}^{\pi} \text{d}\theta^\prime \, f(\vec{r},\theta^\prime,t)   \\
	& \times \int_{-\infty}^{\infty} \text{d}\eta \, p_0(\eta) \,
	\overline{\delta}(\theta^\prime+\eta-\theta) ,
	\end{split}
\end{align}
where the fluctuations are assumed to be distributed according to a Gaussian  $p_0(\eta)$ with a standard deviation $\sigma_0$. The periodicity of angles is accounted for by a sum of $\delta$-functions:  $\overline{\delta}(\theta)=\sum_{m=-\infty}^{\infty}\delta(\theta+2\pi m)$. The parameters $\lambda$ and $\sigma_0$ determine the strength of angular noise in the system. Note that these parameters 
have already been introduced in the multi-particle MD simulations detailed in section~\ref{sect:multi_particle_simulation}~\footnote{Since different kinds of stochastic processes in time can give rise to self-diffusion, we refer to $\lambda$ as a rate within the framework of our Boltzmann approach. However, the periodic stochastic process used in our multi-particle simulations constitutes one of the possible choices.}.
The second term on the r.h.s.\ in \eeqref{eq:BoltzmannEquation} is the collision integral $\mathcal{C}[f^{(2)}]$, which depends on the two-particle density $f^{(2)}$ and captures the effect of binary collisions. It can be split into a \emph{loss}~($-$) and a \emph{gain}~($+$) contribution:
$\mathcal{C}[f^{(2)}]=\mathcal{C}^{-}[f^{(2)}] + \mathcal{C}^{+}[f^{(2)}] $. 
The respective contributions capture the scattering of particles out of or into an angle interval $[\theta,\theta+\text{d}\theta]$ and read~\cite{Bertin_short, Bertin_long}:
\begin{align}\label{eq:collision_integrals}
	&\mathcal{C}^{-}[f^{(2)}] = -  \int_{-\pi}^{\pi} \text{d}\theta^\prime \, \Gamma(\theta^\prime, \theta) f^{(2)}(\vec{r},\theta, \theta^\prime,t) ,\\
	  \label{eq:collision_integral_gain}
	&\mathcal{C}^+[f^{(2)}] =  \int_{-\pi}^{\pi} \text{d}\theta_1 \int_{-\pi}^{\pi} \text{d}\theta_2 \, \Gamma(\theta_1, \theta_2) f^{(2)}(\vec{r},\theta_1,\theta_2,t) \nonumber \\
	 & \quad \times
	 \frac{1}{2} \sum_{j=1}^{2} \int_{-\infty}^{\infty} \text{d}\eta_j \, p_j(\eta_j| \theta_{12}) \, \overline{\delta}(\theta_j+\eta_j-\theta) .
\end{align}

In the \emph{gain} contribution $\mathcal{C}^+[f^{(2)}]$, each of the the two terms ($j=1,2$) accounts for the scattering of one of the particles in a binary collision using the respective distribution $p_1(\eta_1| \theta_{12})$ or $p_2(\eta_2| \theta_{12})$ [\eeqref{eq:scatDist}]. 
The factor $1/2$ is required to avoid counting collisions twice.

The function $\Gamma(\theta_1, \theta_2)$ describes the rate of collisions for a given pre-collisional state determined by the distribution of particles' orientations.
The functional form of $\Gamma(\theta_1, \theta_2)$ can be argued geometrically: 
Consider a collision between two particles with orientations $\theta_1$ and $\theta_2$.
Given short-ranged repulsive interactions, two particles collide if their relative distance becomes less than the particles' diameter $d$. Changing into the reference frame of \emph{e.g.}\ particle $2$, the velocity of particle $1$ is 
given by the relative velocity $\protect{\vec{v}_{12}=v_0\left[ \vec{\hat{v}}(\theta_1) - \vec{\hat{v}}(\theta_2)\right]}$. 
A collision between the two particles occurs within the time interval [$t,t+\text{d}t$]  
if particle $1$ can be found in a rectangle of length $\left|\vec{v}_{12}\right| \text{d}t$ and width $2d$. 
Back in the laboratory frame, this rectangle deforms into a parallelogram retaining its surface area given by
$2dv_0 \left| \vec{\hat{v}}(\theta_1) - \vec{\hat{v}}(\theta_2)\right| \text{d}t=: \Gamma(\theta_1, \theta_2) \text{d}t$ \cite{Bertin_short,Bertin_long}.
This function is commonly referred to as \emph{Boltzmann collision cylinder}. In combination with the two-particle density $f^{(2)}$ it determines the rate of collisions in the pre-collisional state for spherical particles moving ballistically and with constant speed in two dimensions. The function $\Gamma$ only depends on the relative angle $\theta_{12}$, and can be written as  
\begin{equation}\label{eq:boltzmann_cylinder}
	\Gamma(\theta_{12})=4dv_0 \left| \sin{(\theta_{12}/2)}\right|.
\end{equation}
In other systems, the dependence on the relative orientation $\theta_{12}$ may be significantly different, like in the case of ballistically moving rod-shaped particles~\cite{Weber_NJP_2013} where we have $\Gamma(\theta_{12})=4dv_0\left|\sin\left(\theta_{12}/2\right)\right|\left(1+\frac{L/d-1}{2}\left|\sin(\theta_{12})\right|\right)$; here $L$ and $d$ are the lengths of the rods' long and short axis.
In a system of highly diffusive particles like microtubules transported by molecular motors, the dependence on the relative angle may even disappear altogether such that all collisions occur at a constant rate, \emph{i.e.}, $\Gamma(\theta_{12})=\text{const.}$~\cite{Aronson_MT}.

Finally, in order to turn~\eeqref{eq:BoltzmannEquation} into a closed equation for the time evolution of the one-particle density $f$, an expression for the two-particle density $f^{(2)}$ in terms of $f$ has to be postulated. 
In a monoatomic gas, elastic collisions prohibit on average a build-up of inter-particle correlations over time, thereby supporting the validity of the \emph{molecular chaos assumption} [\eeqref{eq:mol_chaos}].
In contrast, in a system consisting of actively propelled constituents, collisions quite generally result in orientational correlations as detailed in section~\ref{sect:scattering_study},
casting doubt on the validity of the molecular chaos assumption.
To account for orientational correlations, we use a modified closure relations for the two-particle density
\begin{equation}\label{eq:no_mol_chaos}
 		f^{(2)}(\vec r, \theta_1,\theta_2,t)= {\chi}(\theta_{12})  f(\vec{r},\theta_1,t) f(\vec{r},\theta_2,t), 
 \end{equation}
\end{subequations} 	
where the function ${\chi}(\theta_{12})$ measures the magnitude of these correlations as a function of the relative angle $\theta_{12}$~\footnote{Note that due to rotational invariance, ${\chi}(\theta_{12})$ solely depends on the relative angle $\theta_{12}$.}. The set of equations~\eqref{eq:BoltzmannEquation_all} represents a generalized kinetic theory for propelled particles moving with constant speed, which is extended compared to Refs.~\cite{Bertin_short, Bertin_long} regarding the following two aspects:  The collision rule is quantitatively determined by the results of the microscopic scattering study, and \eeqref{eq:no_mol_chaos} accounts for angular correlations among the active particles.  

In the following we scrutinize whether these two modifications allow to quantitatively predict the phase boundary obtained from multi-particle MD simulations [see \figref{plot:phase_diagram_simulation}].
To this end, the generalized Boltzmann equation [\eeqref{eq:BoltzmannEquation_all}] is analyzed in terms of Fourier modes, $f(\theta)= (2\pi)^{-1} \sum_{n=-\infty}^{\infty}\fh_n e^{-in\theta}$. Projecting the resulting equation onto the $k$-th Fourier mode $\fh_k$ yields:
\begin{equation}
\label{eq:BoltzmannFourierSpace}
\begin{split}
&\partial_t\fh_k+\frac{v_0}{2}\left[\partial_x(\fh_{k+1}+\fh_{k-1})-i\partial_y(\fh_{k+1}-\fh_{k-1})\right]=\\
&\quad -\lambda \Big( 1-\exp{(-k^2\sigma_0^2/2)}\Big) \fh_k \\
&\quad -\frac{4 g d v_0}{\pi} \sum_{n=-\infty}^{\infty}\fh_n\fh_{k-n} \Big[2\mathcal{I}_n\left[{\chi}\right] -  \mathcal{J}_{n,k}\left[{\chi}\right]\Big] ,
\end{split}
\end{equation}
where the dependence of $\fh_{k}$ on $\vec{r}$ and $t$ was omitted for brevity.
The coefficients in the collision term, $\mathcal{I}_n$ and $\mathcal{J}_{n,k}$, depend on the function ${\chi}(\theta_{12})$ as an additional integration weight and are given by
\begin{align}\label{eq:coeff_I}
	\mathcal{I}_n\left[{\chi}\right] & =\int_{-\pi}^{\pi} \text{d}\theta_{12} \, {\chi}(\theta_{12})  \, \cos{(n\theta_{12})} \, \left|\sin\left({\theta_{12}}/{2}\right)\right| ,\\
\label{eq:coeff_J}
	\mathcal{J}_{n,k}\left[{\chi}\right] & = \mathcal{J}^{(1)}_{n,k}\left[{\chi}\right] + \mathcal{J}^{(2)}_{n,k}\left[{\chi}\right] .
\end{align}
The two contributions to the coefficient $\mathcal{J}_{n,k}$ result from the two terms in the \emph{gain} contribution $\mathcal{C}^+$ of the collision integral \eeqref{eq:collision_integral_gain}. 
These two terms account for the change in the orientation resulting from a collision for particle $1$ and particle $2$, respectively, as denoted by the upper index in brackets. We find
\begin{equation}
\label{eq:coeff_J_integral}
	\mathcal{J}^{(2)}_{n,k}\left[{\chi}\right] = \int_{-\pi}^{\pi} \text{d}\theta_{12} \, {\chi}(\theta_{12}) \, \text{e}^{-in\theta_{12}} G_2(k|\theta_{12}) \, \left|\sin\left({\theta_{12}}/{2}\right)\right| .
\end{equation}
and $\protect{\mathcal{J}^{(1)}_{n,k}= \big(\mathcal{J}^{(2)}_{n,k}\big)^*}$, where we have used that the characteristic functions $G_2(k|\theta_{12})$ and $G_1(k|\theta_{12})$ are related by complex conjugation [\eeqref{eq:charfkt_cc}].

\begin{figure*}[htp]
   \centering
\begin{tikzpicture}
    \node[anchor=south west,inner sep=0] (image) at (0,0) {\includegraphics[width=\linewidth]{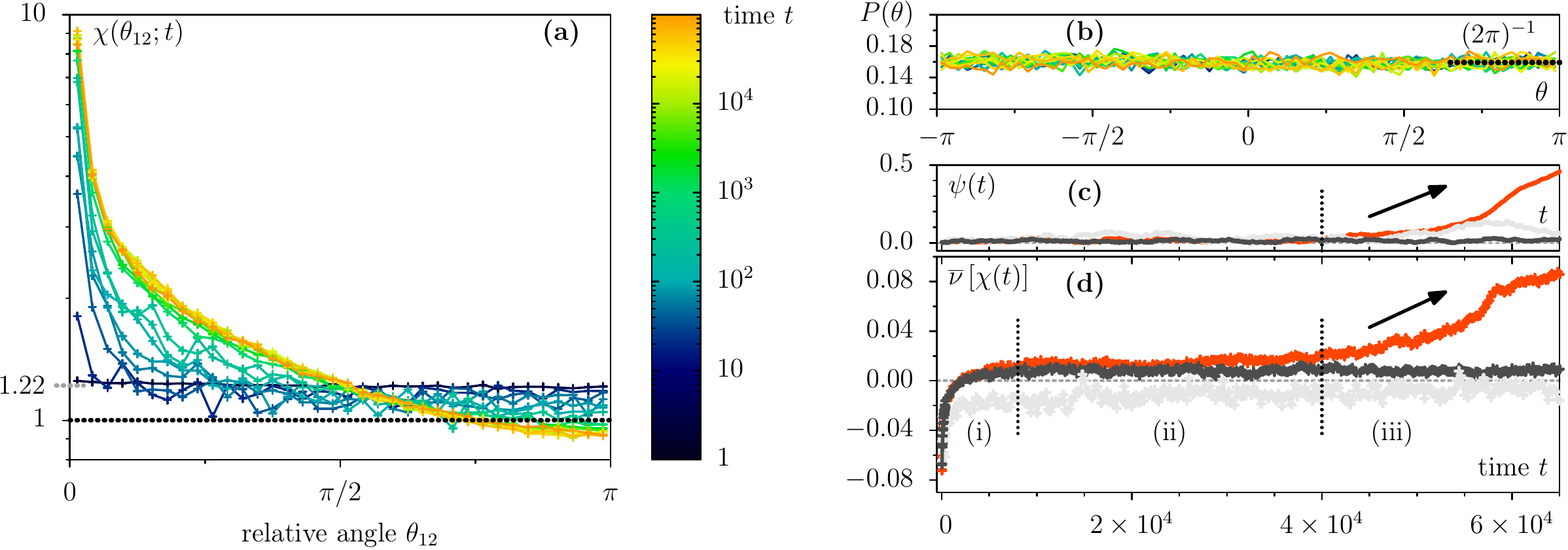}};
    \begin{scope}[x={(image.south east)},y={(image.north west)}]
	\node[anchor=south west,inner sep=0] (image_inset) at (0.20,0.43) {\includegraphics[width=0.16\linewidth]{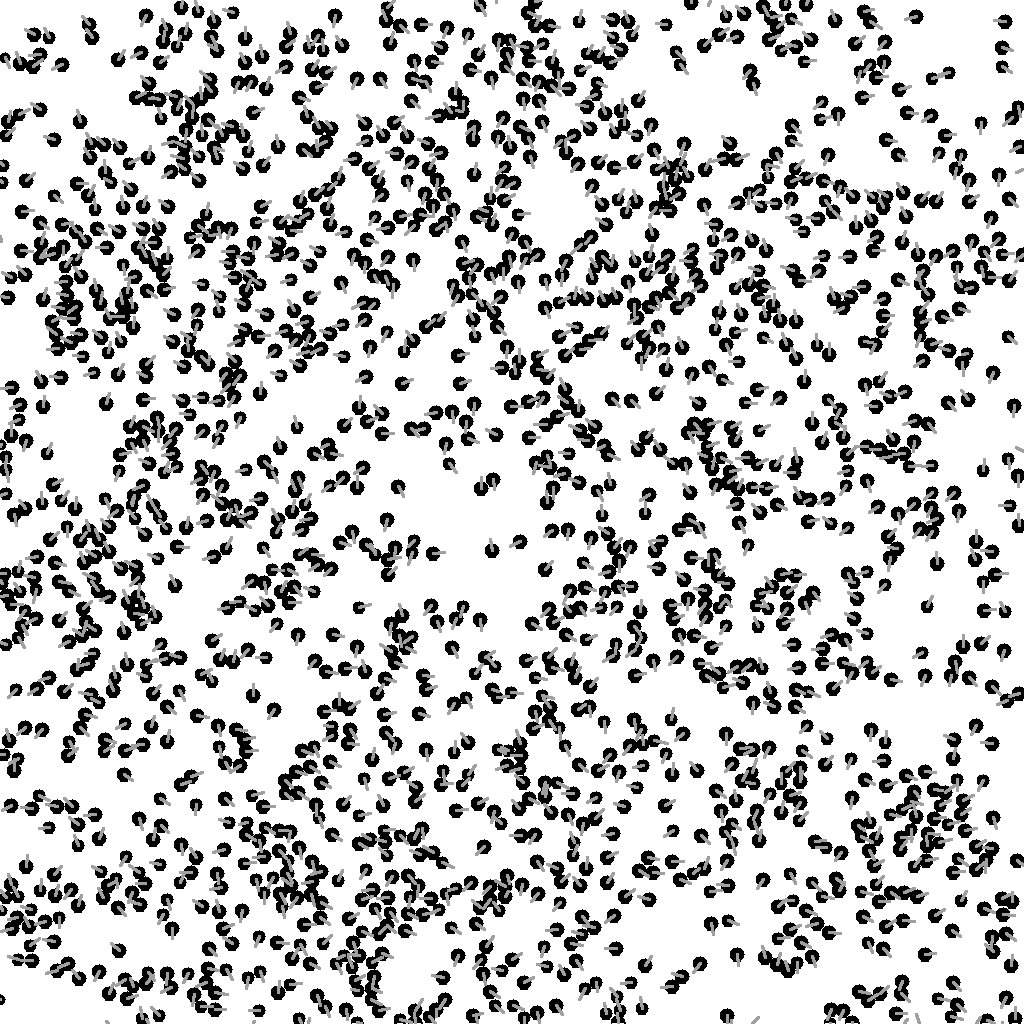}};
	\draw[thick,color=black] (image_inset.south west) rectangle (image_inset.north east);
    \end{scope}
\end{tikzpicture}
\caption{
(color online)
\textbf{(a)} \emph{Distribution ${\chi}(\theta_{12};t)$} as a function of pre-collisional relative angle $\theta_{12}$ for a control parameter set $(\Phi, \sigma_0)=(0.2,0.42^{\circ})$ close to phase boundary [see \figref{plot:phase_diagram_simulation}] on the isotropic side. 
The initial constant distribution (black) evolves to a distribution favoring smaller relative angles (lighter shading) proving the emergence of  angular correlations \CH{(``precursor correlations")}. 
${\chi}(\theta_{12};t)$ is measured at different times $t$ within a sampling time interval $[t-\tau,t]$ of length $\tau$ [see \eeqref{eq:coll_rate_V_simplified}].
Time points $t$ are indicated by the shading; lengths of the sampling intervals are: $\tau=20$ for $t\in[20,400]$, and $\tau=200$ for $t\in[600,10^5]$ with time $t$ given in units of $\tau_\text{eq}$. 
\emph{Inset:} Snapshot of a small section ($75d\times75d$) of the system at time $t=2\times 10^4$ showing a mostly homogeneously distributed particle density.
\textbf{(b)} \emph{Probability distribution $P(\theta)$ of particles' orientations} at time-points $t$ equal to those shown in \textbf{(a)}. Over the whole time $t\in[0,10^5]$, there is no discernible deviation from the isotropic state $P(\theta)=(2\pi)^{-1}$.
\textbf{(c)} \emph{Polarization $\psi(t)$} and  \textbf{(d)}  $\overline{\nu}\left[{\chi}(t)\right]$ as a function of time for an isotropic state [dark grey, $(\Phi, \sigma_0)=(0.2,0.42^{\circ})$] and polar ordering state [red (medium grey), $(\Phi, \sigma_0)=(0.2,0.38^{\circ})$]. 
For both parameter sets, $\overline{\nu}\left[{\chi}(t)\right]$ rapidly increases (i) from an initial negative value (corresponding to ${\chi}_\text{init}(\theta_{12})=\text{const.}$), 
and becomes positive. Then, $\overline{\nu}\left[{\chi}(t)\right]$ exhibits for both cases a prolonged plateau (ii) with $\overline{\nu}\approx0.008$ for $\sigma_0=0.42^{\circ}$ and $\overline{\nu}\approx0.014$ for $\sigma_0=0.38^{\circ}$, corresponding to correlated states [see \textbf{(a)}]. 
However, for times $t<4\times10^4$, $\psi(t)\approx0$ for both states.
For the polar ordering state the transition occurs at approximately $t\approx4\times 10^4$, reflected in $\psi\to1$ in \textbf{(c)} and a strong increase (iii) away from the plateau of $\overline{\nu}\left[{\chi}(t)\right]$.
Measurements were performed in a system of linear size $500d$.
\CH{Additionally, an isotropic [$\psi(t)\approx0$ in \textbf{(c)}] parameter set for a lower packing fraction [light grey, $(\Phi, \sigma_0)=(0.1,0.21^{\circ})$] is shown, where $\overline{\nu}\left[{\chi}(t)\right]$ remains negative most of the time, but intermittently peaks at positive values close to the plateau found for $\Phi=0.2$ [dark grey line in \textbf{(d)}]. 
It is known that the close to the phase boundary
the nucleation of a cluster of sufficiently large mass
triggers the transition to collective motion~\cite{Weber_nucleation}, therefore we interpret
 $\max_t \overline{\nu}\left[{\chi}(t)\right]>0$ as the results of precursor correlations which are necessary for the ensuing transition described by the continuous Boltzmann Equation. 
}}
\label{plot:RelativeAngleDistribution}
\end{figure*}

Using the Boltzmann equation in Fourier space [\eeqref{eq:BoltzmannFourierSpace}] as a starting point, one can determine the onset of polar order in a homogeneous system as follows. The first two Fourier components determine the (hydrodynamic) particle density $\rho=\fh_0$ and the momentum density $\vec{g}=\rho\vec{u}=v_0\fh_1$, where $\vec{u}$ denotes the hydrodynamic velocity. Since $\rho$ is conserved and the momentum field $\vec{g}$ plays the role of the broken symmetry variable, both fields are slow hydrodynamic variables.
Near the onset of the instability of the isotropic state, $f(\vec{r},\theta,t)=\rho_0/(2\pi)$, the hydrodynamic velocity $|\vec u|$ is small compared  to the microscopic driving velocity $v_0$. Since $\fh_k=\mathcal{O}\left[(|\vec u|/v_0)^k\right]\ll1$, we are thus able to truncate \eeqref{eq:BoltzmannFourierSpace} by setting $\fh_k\approx0$ for all $ k>2$ \cite{Aronson_MT, Bertin_short, Bertin_long}. Since we are only interested in the location of the phase boundary marking the transition from the homogeneous isotropic to the polarized state, we neglect all spatial gradients $\mathcal{O}(\nabla \vec g)$, $\mathcal{O}(\nabla \rho)$ and $\mathcal{O}(\vec g^3)$ \footnote{Including higher order terms or use of a different truncation scheme \cite{Peshkov:2012tu} does not affect the coefficient $\nu$ of  the linear term.}, and find the following set of equations: 
\begin{subequations}
\label{eq:HydrodynamicEquations}
\begin{align}
\label{eq:HydrodynamicEquationsRho}
\partial_t \rho&=0,\\
\label{eq:HydrodynamicEquations_g}
\partial_t\vec{g}&=\nu \, \vec{g}.
\end{align}
\end{subequations}
The kinetic coefficient $\nu$ determines the linear stability of the initially isotropic state: for $\nu>0$ polar order may develop, whereas $\nu<0$ signals that the isotropic state is linearly stable. 
We find
\begin{equation}\label{eq:coeff_nu}
	\nu(\Phi,\sigma_0) = -\lambda \big( 1-\text{e}^{-\sigma_0^2/2} \big) + \overline{\nu}\left[{\chi}\right]  \frac{4 v_0}{\pi^2d}  \Phi  ,
\end{equation}
where the packing fraction $\Phi=\rho_0\pi d^2/4$ and 
$\overline{\nu}\left[{\chi}\right]$ reads in terms of the scattering coefficients defined in \eeqref{eq:coeff_I} and \eeqref{eq:coeff_J}:
\begin{equation}\label{eq:coeff_nu_bar}
	\overline{\nu}\left[{\chi}\right] = \mathcal{J}_{0,1}\left[{\chi}\right] + \mathcal{J}_{1,1}\left[{\chi}\right] - 2\mathcal{I}_0\left[{\chi}\right] - 2\mathcal{I}_1\left[{\chi}\right] .
\end{equation}
The function ${\chi}$, which describes angular correlations, determines the sign and magnitude of $\overline{\nu}\left[{\chi}\right]$, and is thereby a main determinant of the phase boundary between the isotropic and polarized states. Note that for $\overline{\nu}\left[{\chi}\right]<0$, the isotropic state is linearly stable for all values of the control parameters $(\Phi,\sigma_0)$.

\subsection{Phase boundary}\label{sect:collective_comparison}

The condition $\nu(\Phi, \sigma_0)=0$ determines the phase boundary as a function of the control parameters, \emph{i.e.}, single particle noise strength $\sigma_0$ and packing fraction $\Phi$.
Solving for the critical single particle noise strength $\sigma_{0,\text{c}}(\Phi)$ as a function of the packing fraction, we find:
\begin{equation}\label{eq:sigma_c}
	\sigma_{0,\text{c}}(\Phi)[{\chi}]=\bigg[ {-2} \ln \bigg(1 - \overline{\nu}\left[{\chi}\right] \frac{4 v_0}{\pi^2d\lambda}\Phi\bigg)\bigg]^{1/2}.
\end{equation}
Above this threshold noise strength $\sigma_{0,c}$ the isotropic state is stable, whereas for $\sigma_0 \le \sigma_{0,c}$ a macroscopic polarized state can develop. For the specific parameters used in the MD simulations in section~\ref{sect:multi_particle_simulation} one has $v_0(d\lambda)^{-1}=\mu(\lambda\tau_{eq})^{-1}=0.005$.

\subsubsection*{Specification of the analytical phase boundary} \label{sect:collective_specification}

At this point, we have to specify the function ${\chi}(\theta_{12})$ in \eeqref{eq:sigma_c} to compute the  phase boundary, and compare it to the one obtained from multi-particle MD simulations [see \figref{plot:phase_diagram_simulation} and section~\ref{sect:multi_particle_simulation} for corresponding discussion].

Let us first calculate the  phase boundary by assuming that the initial states at the onset of collective motion are devoid of angular correlations, \emph{i.e.}, the \emph{assumption of molecular chaos} is fulfilled with ${\chi}(\theta_{12})=1$ in~\eeqref{eq:no_mol_chaos}.
In this case we find that  $\overline{\nu}\left[{\chi}=1\right]<0$,
implying that $\nu<0$ for all control parameters $\Phi$ and $\sigma_0$ [see \eeqref{eq:coeff_nu}]. 
Hence, one would conclude that the system's isotropic state remains stable for arbitrary values of the control parameters, which is obviously at odds with the phase diagram obtained from multi-particle simulations [\figref{plot:phase_diagram_simulation}]. 
This clearly indicates that the state of the system preceding a transition to a polarized state cannot be free of angular correlations.

To further scrutinize this finding, we ran multi-particle MD simulations starting from an initially uncorrelated, homogeneous and isotropic state and studied ${\chi}(\theta_{12};t)$ as a function of time.
For this purpose we selected a set of control parameters ($\Phi$,$\sigma_0$) sufficiently close to the phase boundary obtained from MD simulations for which the system remains isotropic.
The function ${\chi}(\theta_{12};t)$ was computed  by recording the relative angles $\theta_{12}$ of all collisions occurring in the simulation box of area $V=L^2$ within a sampling time interval $[t-\tau,t]$. 
The length of the sampling time interval $\tau$ determines the number of recorded collisions, yet has to be chosen small enough to properly resolve the time evolution of ${\chi}(\theta_{12};t)$ [values of  $\tau$ are given in the caption of \figref{plot:RelativeAngleDistribution}].
Noting that $ -\frac{1}{2}\int_V\text{d}^2r\int_{-\pi}^{\pi}\text{d}\theta \, \mathcal{C}^{-}[f^{(2)}]$, with $\mathcal{C}^{-}[f^{(2)}]$ given in \eeqref{eq:collision_integrals}, is equal to the total rate of binary collisions in a volume $V$, the collision frequency 
$\omega_V(\theta_{12};t)$ as a function of the relative angle $\theta_{12}$ can be written as
\begin{equation}\label{eq:coll_rate_V}
\begin{split}
	\omega_V(\theta_{12}; t) = & \frac{1}{2} \int_V\text{d}^2r \int_{-\pi}^{\pi}\text{d}\theta_1^\prime \int_{-\pi}^{\pi}\text{d}\theta_2^\prime \, \delta(\theta_{12}^\prime-\theta_{12}) \\
	& \times \Gamma(\theta_{12}^\prime)  \, {\chi}(\theta_{12}^\prime;t) \, f(\vec r, \theta_1^\prime,t) f(\vec r, \theta_2^\prime,t)  .
\end{split}
\end{equation}
The above relation connects 
 ${\chi}(\theta_{12};t)$ with the collision frequency 
$\omega_V(\theta_{12};t)$, which can be measured in the multi-particle simulations.
In our subsequent numerical studies we only consider states that remain approximately homogeneous and isotropic in time. Therefore, we assume  
  $f=\rho_0/(2\pi)$, which allows to considerably simplify the above equation to: 
\begin{equation} \label{eq:coll_rate_V_simplified}
	{\chi}(\theta_{12};t) = \frac{4\pi}{\Gamma(\theta_{12}) \,  \rho_0^2 V} \, \omega_V(\theta_{12};t).
\end{equation}
The initial configuration in the MD simulations was a homogenous and isotropic state. To this end, the initial positions of the particles and the orientations of their velocities were chosen randomly, which was followed by a relocation of overlapping particles until excluded volume had been enforced for all particles. In this initial state, angular correlations are absent and therefore ${\chi}_\text{init}(\theta_{12})=\text{constant}$.
The assumption of molecular chaos implies ${\chi}(\theta_{12})=1$, however the initial value in our system is expected to be larger. This is due to the finite size of our active spheres, which reduces the amount of free-volume in the system and in turn gives rise to spatial correlations between the particles.
In kinetic theory for hard granular gases this is accounted for by the so called Enskog factor $g(\Phi)$~\cite{Brilliantov_book}. It depends on the packing fraction $\Phi$, and is given in two dimensions by the following approximation~\cite{verlet_levesque_1982}:
\begin{equation} \label{eq:enskog_2d}
	g(\Phi) =\frac{1-\frac{7}{16}\Phi}{(1-\Phi)^2}.
\end{equation}
Measured in our system of active particles for a control parameter set ($\Phi$,$\sigma_0$) where the system remains isotropic, \figref{plot:RelativeAngleDistribution}(a) depicts the time evolution of ${\chi}(\theta_{12};t)$.
The initial distribution of ${\chi}(\theta_{12})$ [\figref{plot:RelativeAngleDistribution}(a), black curve] is indeed shifted to a numerical value slightly larger than $1$, reflecting the presence of spatial correlations. However, for the packing fraction $\Phi=0.2$ used in~\figref{plot:RelativeAngleDistribution}, the Enskog factor [\eeqref{eq:enskog_2d}] is $g\approx 1.4$ and therefore slightly overestimates the actual increase of ${\chi}(\theta_{12})$ found in the MD simulations, \CH{which is approximately equal to ${\chi}_\text{init}\approx1.22$}.  We attribute this small discrepancy to the fact that our active colloids are soft, consistently leading to a smaller amount of decreased free-volume, and thereby a lower value of $g$.

As time progresses, we find that ${\chi}(\theta_{12};t)$ evolves to a distribution favoring smaller relative angles [\figref{plot:RelativeAngleDistribution}(a)], while the system remains isotropic and uniform. The latter is reflected  in a flat angular probability  distribution $P(\theta)$ [\figref{plot:RelativeAngleDistribution}(b)], and a polarization $\psi\approx0$ (\figref{plot:RelativeAngleDistribution}(c), dark grey curve; see also video in Supplemental Material~\cite{Supplement_Videos}).
The deviation from a uniform ${\chi}(\theta_{12})$  clearly indicates that angular correlations in the system develop as the system approaches its stationary state.
Concomitantly, the coefficient $\overline{\nu}$ shows (i) a rapid increase from an initial negative value $\overline{\nu}_\text{init}\approx-0.07$ (corresponding to ${\chi}_\text{init}\approx1.2$) to 
(ii) a prolonged plateau at a positive value $\overline{\nu}\approx0.008$ [\figref{plot:RelativeAngleDistribution}(d), dark grey curve]. 
This sign change in $\overline{\nu}$ is triggered by the emerging angular correlations and allows the generalized Boltzmann approach to predict an ordering transition.

A quantitatively similar behavior of $\overline{\nu}$ as a function of time can be found for control parameter sets ($\Phi$,$\sigma_0$) close to the phase boundary that give rise to a polar ordering transition [red dots in \figref{plot:phase_diagram_simulation}]. 
The only difference manifests in a subsequent third regime (iii) where the system  begins to develop a polarized state with $\psi\to1$, which is reflected by a further increase of $\overline{\nu}\left[{\chi}(t)\right]$ away from the plateau value.
The prolonged plateau can be interpreted as a \emph{lag phase}, in which the system remains isotropic [$\protect{\psi\approx0}$, \figref{plot:RelativeAngleDistribution}(c), red (light grey) curve] and ``waits'' for the nucleation of a cluster of sufficiently large size~\cite{Weber_nucleation}.
Please refer to the Supplemental Material~\cite{Supplement_Videos} for a video depicting the time evolution of the system from the initial configuration to the fully polarized state.
Taken together, close to the phase boundary---on the isotropic as well as the polar side---orientational correlations exist which are the essential prerequisites for a subsequent transition to a polar state. These correlations are a \emph{precursor} phenomenon that precedes the phase transition.

\begin{figure}[tbp]
\begin{center}
\includegraphics[width=\linewidth]{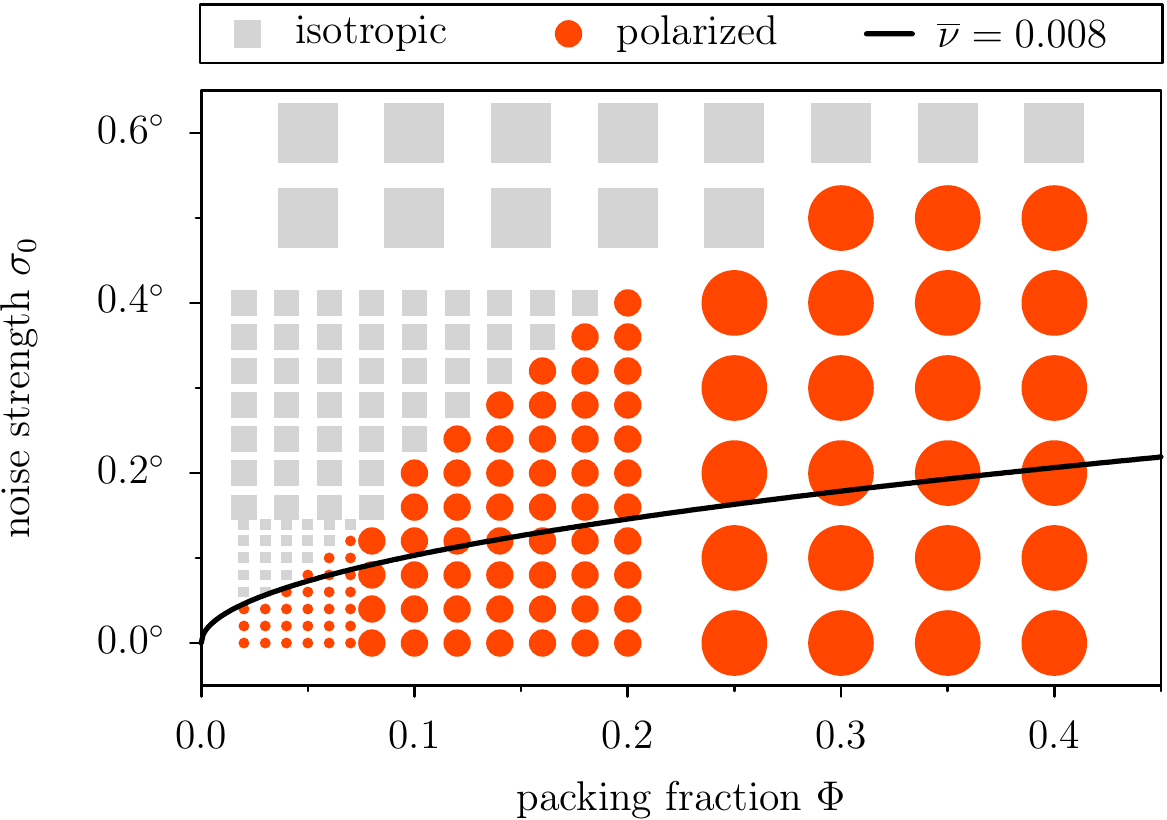}
\caption{
(color online)
\emph{Phase diagram} as function of packing fraction $\Phi$ and angular noise $\sigma_0$ [degrees]: 
Results obtained from multi-particle simulations are displayed by symbols, while the solid line corresponds to the predictions of kinetic theory for the phase boundary. 
\emph{Multi-particle simulation:}
Red dots indicate polarized states while grey squares correspond to control parameter sets where no change in the initial isotropic state has been observed. 
\emph{Analytical results:} 
The solid line indicates the phase boundary obtained from kinetic theory [\eeqref{eq:sigma_c}] using ${\chi}(\Phi=0.2)$ [resulting in $\overline{\nu}=0.008$; see \figref{plot:RelativeAngleDistribution}(d)] for all $\Phi$ and $\sigma_0$. 
}
\label{plot:phase_boundary}
\end{center}
\end{figure}

Now we would like to study the implications of these \emph{precursor} correlations on the phase boundary [\eeqref{eq:sigma_c}].
To this end, we use the plateau value of $\overline{\nu}\approx 0.008$ [see \figref{plot:RelativeAngleDistribution}(d), dark grey curve],
and assume that the underlying ${\chi}$ is valid for all packing fractions $\Phi$ and noise strengths $\sigma_0$.
The corresponding result for the phase boundary $\sigma_{0,\text{c}}(\Phi)$ is depicted by the solid line in~\figref{plot:phase_boundary}. 
It nicely agrees with the phase boundary obtained from multi-particle simulations for small packing fractions. 
This indicates that our extended kinetic theory for propelled particle systems constitutes a quantitative~\CH{\footnote{\CH{Preliminary numerical solutions~\cite{Thuroff_2013} to the Boltzmann equation \eeqref{eq:BoltzmannEquation_all} indicate that the homogenous phase boundary predicted by 
\eeqref{eq:sigma_c} also constitutes the \emph{quantitatively} correct 
phase boundary for the inhomogeneous Boltzmann equation
when starting from an initial disordered state.}}}
 description for soft active colloids.
Further our findings stress the significance of correlations in active systems at the onset of collective motion.
However, for larger packing fractions, polar order persists beyond the critical noise strength predicted by the Boltzmann theory. This increased stability of polar order with respect to noise may be attributed to clustering processes in the regime of intermediate packing fractions. How such effects are properly accounted for within a kinetic theory is presently unclear. One will certainly need to go beyond \eeqref{eq:no_mol_chaos} and account for higher-order correlations, or employ a multi-species formulation~\cite{Weber_NJP_2013}.

\section{Summary \& Outlook}\label{sect:conclusion_outlook}

In this paper we addressed the question---focusing on dilute conditions---whether we can understand the collective dynamics of soft active colloids by  solely considering \emph{binary} interactions between the constituent particles. 
To this end we performed binary scattering studies of active colloids interacting via soft repulsive interactions. 
From these studies we discovered that the dynamic models considered share the same principle of parallel alignment. The underlying principle was termed as ``\emph{alignment of the first},'' stating that the first incoming particle with respect to the center of collision is aligned to the second of the colliding particles. 
We showed that these types of collisions contribute the dominant part to the system's alignment tendency.
Additionally, this principle is genuinely different to alignment in systems of inelastic gases, which is mediated by damping of the relative particle velocities.
Moreover, from the binary scattering studies we deduced a non-linear collision rule mapping  pre-collisional on post-collisional velocities that is devoid of any approximation.
This collision rule was then connected to the system's collective behavior by kinetic theory for propelled particle systems. 
The microscopic origin of the collision rule allowed to quantitatively scrutinize the predictions of kinetic theory with regard to the phase boundary marking the instability of the isotropic unpolarized state.
By comparing the resulting phase boundary with that obtained from multi-particle simulations of the underlying microscopic model for active colloids, we discovered that non-trivial modifications in the kinetic description are necessary to obtain a quantitative agreement in the phase boundary. 
Specifically, we found that \emph{precursor} orientational and spatial correlations exist
close to the phase boundary.  
Only if the kinetic description included these correlations, the analytic prediction for the phase boundary coincided quantitatively for small packing fractions with the one from multi-particle simulations.
Most importantly, if orientational correlations were neglected, kinetic theory for propelled particles failed, \emph{i.e.}, it predicted that ordering is absent, which is at odds with corresponding molecular dynamics simulations.

Our findings clearly indicate that the framework of kinetic theory for propelled particle systems is flexible enough to accommodate the complex behavior of soft active colloids 
 and allow a bottom-up understanding of how the microscopic dynamics of binary collisions
 is related to the system's behavior on large length and time-scales.
 The developed ``renormalized" kinetic theory, where the interaction kernel, \emph{i.e.},  the collision rule and the correlations of the pre-collisional state, are determined from microscopic molecular dynamics simulations,
 could serve as the appropriate starting point for an extension of kinetic theory for propelled particle systems into the regime of intermediate packing fractions.   
Moreover, we are convinced that our approach is also perfectly suited to bridge between microscopic experimental studies of propelled particle systems~\cite{Schaller,Yutaka,Dauchot_long,VibratedDisks2013}, in which \emph{precursor} correlations are likely to exist, and their corresponding quantitative mesoscopic description.  

\begin{acknowledgments}
The authors would like to thank Florian Th\"uroff  for discussions and critical reading of our manuscript. 
This project was supported by the Deutsche Forschungsgemeinschaft in the framework of the SFB 863 and the German Excellence Initiative via the program ``Nanosystems Initiative Munich'' (NIM).
\end{acknowledgments}


\appendix

\section{Derivation of the alignment integral} 
\label{appendix:alignment_integral}

\begin{figure}[tb]
\begin{center}
\begin{tikzpicture}
  \usetikzlibrary{decorations.pathreplacing}
  \begin{scope}[scale=0.8, dashedlines/.style={thick,color=gray,densely dashed}, lines/.style={thick,color=black}, circles/.style={thick,color=black}, arrow/.style={lines,very thick,->,>=latex},]
	\draw[circles] (3.8,1) node[coordinate] (p) {} +(10:1) arc(10:350:1);
	\draw[dashedlines] (p) +(10:2) arc(10:350:2);
	\path (p) ++(10:2) node[coordinate] (cylinder_start_left) {} +(30:3) node[coordinate] (cylinder_end_left) {};
	\path (p) ++(0:2) node[coordinate] (cylinder_start_middle) {} +(30:2.5) node[coordinate] (cylinder_nearend_middle) {};
	\path (p) ++(-10:2) node[coordinate] (cylinder_start_right) {} +(30:3) node[coordinate] (cylinder_end_right) {};
	\filldraw[dashedlines,fill=gray!50] (p) -- (cylinder_start_left) -- (cylinder_start_right) -- cycle;
	\draw[lines] (cylinder_start_left) -- (cylinder_end_left) -- (cylinder_end_right) -- (cylinder_start_right) -- cycle;
	\draw[arrow] (p) -- node[above=2, pos=0.3] {$\vec{\hat{e}}$} +(0:1.5);
	\draw[dashedlines] (cylinder_nearend_middle) -- (cylinder_start_middle);
	\draw[arrow] (cylinder_nearend_middle) -- node[fill=white,pos=0.4,sloped] {$\vec{v}_{12}$} +(30:-1.9);
	\draw[dashedlines] (cylinder_end_right) -- ++(0,-1.5) +(0,-0.15) coordinate (cylinder_brace_start);
	\path (cylinder_start_right) +(0,-0.15) coordinate (cylinder_brace_end);
	\draw[lines,decorate,decoration={brace}] (cylinder_brace_start) -- node[pos=0.45,below=3pt] {$\left| \vec{v}_{12} \cdot \vec{\hat{e}} \right| \varDelta t$} (cylinder_brace_end);
	\path (p) ++(0.6,1.9) node[fill=white] (text_1) {$d \, \text{d}\vec{\hat{e}}$};
	\draw[thick,color=black!70,>=latex,<-,shorten <= 2pt] (cylinder_start_left) +(0,-0.25) -- (text_1.south);
	\draw[lines,|-|] (p) ++(1,-1.35) -- node[fill=white,pos=0.5] {$d$} +(180:2);
  \end{scope}
\end{tikzpicture}
\caption{Illustration of the collision cylinder. Within a time interval $\varDelta t$, a second particle collides with the shown particle with the point of contact given by the collision vector $\vec{\hat{e}}$ if its center lies in the collision cylinder with volume $\text{d}V_\text{cc} = d \, \text{d}\vec{\hat{e}} \, \left| \vec{v}_{12} \cdot \vec{\hat{e}} \right| \varDelta t$, where $d$ is the particles' diameter and $\vec{v}_{12}$ the relative velocity.}
\label{pic:collision_cylinder}
\end{center}
\end{figure}
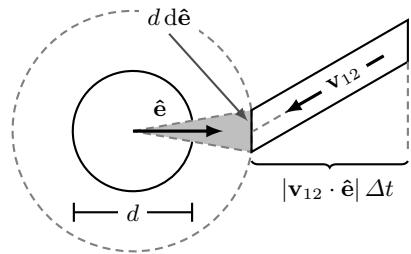

In the following we derive the integration weight used in \eeqref{eq:alignment_integral}, which depends on the specific collision geometry. 
To this end we assume uniformly distributed positions and velocities and consider a collision of particles with diameter $d$, relative velocity $\vec{v}_{12}$, and the point of contact defined by the unit vector $\vec{\hat{e}}$ [see \figref{pic:coll_geometry}]. 
The likelihood of such a collision is proportional to the number of particles in the Boltzmann collision cylinder [\figref{pic:collision_cylinder}] with volume
\begin{equation}
  \text{d}V_\text{cc} \left(\vec{\hat{e}},\vec{v}_{12}\right) = d\,  \left| \vec{v}_{12} \cdot \vec{\hat{e}} \right| \text{d}\vec{\hat{e}} \, \varDelta t .
\end{equation} 
The number of particles with velocity $\vec{v}_1$ in the collision cylinder is proportional to $\protect{f \left( \vec{v}_1 \right) \, \text{d}\vec{v}_1 \, \text{d}V_\text{cc}}$, where $f \left( \vec{v}_1 \right)$ is the one-particle distribution function, which is independent of position in a homogeneous system.
Integration yields the total number of particles that collide with a given particle with velocity $\vec{v}_2$ per time interval $\varDelta t$: 
\begin{equation} \label{eq:general_rate_of_collisions}
  {\varDelta t}^{-1} \int \int f \left( \vec{v}_1 \right) \, \Theta \left( - \vec{v}_{12} \cdot \vec{\hat{e}} \right) \, \text{d}\vec{v}_1 \, \text{d}V_\text{cc} ,
\end{equation}
where the Heaviside function ensures that the particles collide [see \figref{pic:collision_cylinder}]. To rewrite \eeqref{eq:general_rate_of_collisions} as an integral over the impact parameter $b$, we employ the angle $\gamma$ [defined in \figref{pic:coll_geometry}(b)] as an intermediate step:
\begin{equation} \label{eq:rewrite_de_to_db_1}
  \int \text{d}\vec{\hat{e}} \ \Theta \left( - \vec{v}_{12} \cdot \vec{\hat{e}} \right) \left| \vec{v}_{12} \cdot \vec{\hat{e}} \right| \, = \, \left| \vec{v}_{12} \right| \int_{0}^{\pi} \text{d}\gamma \, \sin\gamma ,
\end{equation}
where we used $\left| \vec{v}_{12} \cdot \vec{\hat{e}} \right| = \left| \vec{v}_{12} \right| \, \sin\gamma$. Using the definition of the impact parameter as $b = - \cos\gamma$ [\eeqref{eq:def_impact}], we arrive at
\begin{equation} \label{eq:rewrite_de_to_db_2}
  \left| \vec{v}_{12} \right| \int_{0}^{\pi} \text{d}\gamma \, \sin\gamma \, = \, \left| \vec{v}_{12} \right| \int_{-1}^{+1} \text{d}b \text{.}
\end{equation}
Since there is no weight function depending on $b$ left in the integral, collisions with different impact parameters occur at the same rate provided that the system is homogeneous and devoid of correlations. Note that this result is valid independent of the distribution of particle velocities.
For the system considered here, we make two assumptions concerning the velocity distribution function: (i)~Particle velocities are relaxed to the stationary value $v_0$ before a collision, \emph{i.e.}, $f(\vec{v}_1)=\delta(v_0-|\vec{v}_{1}|)f(\theta_{12})$ where the relative angle $\theta_{12}$ denotes the direction of $\vec{v}_1$ with $\vec{v}_2$ defining the reference axis. (ii)~Directions of particle velocities are isotropically distributed, \emph{i.e.}, $f(\theta_{12}) = \text{constant}$.
Taken together, this yields for the rate of collision given in \eeqref{eq:general_rate_of_collisions}
\begin{equation}
  \text{const.}\times \int \text{d}\theta_{12}  \int_{-1}^{+1} \text{d}b \, \left| \sin \left(\frac{\theta_{12}}{2}\right) \right| ,
\end{equation}
where we used that the norm of the relative velocity $|\vec{v}_{12}|=\CH{2}v_0|\sin(\theta_{12}/2)|$. The intuitive interpretation for the weight function is that two particles moving in the same direction with equal speed will never collide, whereas particles moving in opposite directions have maximal relative velocity and therefore occur most often.
Particle exchange symmetry allows to restrict the range of integration to $\theta_{12} \in [0,\pi]$. Therefore,
 we can define the following normalized weighted average over all collision geometries:
\begin{equation}\label{eq:alignment_integral2}
  \langle ... \rangle \, = \, \frac{1}{4} \int_{-1}^{+1} \text{d}b \int_{0}^{\pi} \text{d}\theta_{12} \, \left( ... \right) \, \left| \sin\left(\frac{\theta_{12}}{2}\right) \right| .
\end{equation}
Using \eeqref{eq:alignment_integral2} for the relative alignment $\Delta A$ [\eeqref{eq:def_relative_alignment}], we arrive at the alignment integral given in \eeqref{eq:alignment_integral}.

\section{Gaussian approximation for the scattering distribution} 
\label{appendix:gaussian_approximation}

\begin{figure}[b]
\begin{center}
\includegraphics[width=0.49\linewidth]{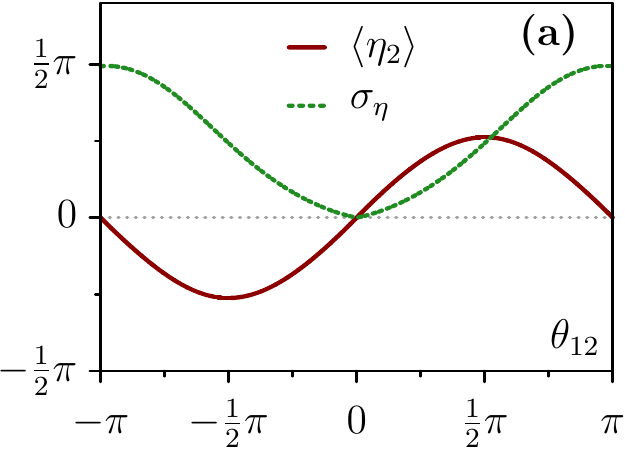}
\hfill
\includegraphics[width=0.49\linewidth]{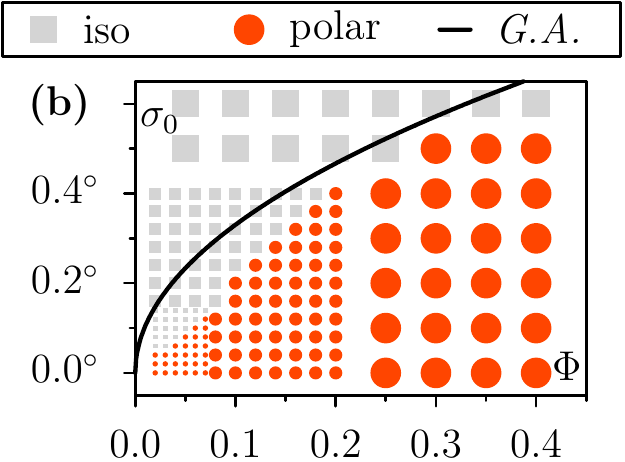}
\caption{
(color online)
\textbf{(a)} Mean $\langle\eta_2\rangle$ (solid red line) and standard deviation $\sigma_\eta$ (dashed green line) of the scattering distribution $p_2(\eta_2|\theta_{12})$ [cf.~\figref{plot:scatDist_charFunct}]. 
Since particles are identical, the corresponding deviations are equal and $\langle\eta_1\rangle=-\langle\eta_2\rangle$.
\textbf{(b)} \emph{Phase diagram} as a function of packing fraction $\Phi$ and angular noise $\sigma_0$ [degrees], as described in caption of Fig.~\ref{plot:phase_boundary}. 
The solid line corresponds to the phase boundary obtained from the kinetic approach using a \emph{Gaussian approximation} (\emph{G.A.}) for the scattering distribution and assuming  \emph{molecular chaos}, \emph{i.e.}, ${\chi}=1$.}
\label{plot:gaussian_approximation}
\end{center}
\end{figure}

In order to quantitatively scrutinize the predictions for the phase boundary obtained from kinetic theory, accounting for
the full characteristic function [\eeqref{eq:def_charfkt}] is crucial.  
This can be seen when considering a \emph{Gaussian approximation} for the scattering distribution $p_2(\eta_2|\theta_{12})$ [cf.~\figref{plot:scatDist_charFunct}], \emph{i.e.}, expressing the characteristic function through the moments of the distribution
 and keeping only the first two terms
\begin{equation} \label{eq:gauss_approx_charFunct}
	G_j^{\text{Gauss}}(k|\theta_{12}) = e^{ik\langle\eta_j\rangle(\theta_{12})} \, e^{-k^2 \sigma_{\eta}^2(\theta_{12})/2},
\end{equation}
where
 $\langle\eta_2\rangle(\theta_{12})$ denotes the mean scattering angle and  $\sigma_{\eta}(\theta_{12})$ is the standard deviation of the scattering distribution; both are depicted in \figref{plot:gaussian_approximation}(a).
Using ${\chi}=1$ (\emph{assumption of molecular chaos}) and $G_j^{\text{Gauss}}$ to compute the scattering coefficients $\mathcal{J}_{n,k}$ [\eeqref{eq:coeff_J_integral}], the phase boundary in the \emph{Gaussian approximation} can be determined; see solid black line in~\figref{plot:gaussian_approximation}(b).
In section~\ref{sect:collective_comparison} we showed that \emph{precursor} 
correlations exist at the onset of collective motion, and are essential to reproduce quantitatively the phase boundary obtained from MD simulations. 
Even though the \emph{Gaussian approximation} correctly predicts the existence of an ordering transition, this is only due to an inaccurate approximation:   
It misrepresents the actual scattering behavior by underestimating the impact of scattering events with large relative angles, and thus masks the true reason for the instability.


%

\end{document}